\documentclass[reprint, aps,amssymb,amsmath,showpacs] {revtex4-2}%

\usepackage{graphicx}
\usepackage{dcolumn}
\usepackage{bm}
\usepackage{siunitx}%
\usepackage[version=4]{mhchem}
\usepackage{xspace}
\usepackage{booktabs}
\usepackage{float}
\usepackage{placeins}
\usepackage{lineno}
\usepackage{subfigure}
\usepackage{enumerate}
\usepackage{hyperref}
\usepackage{enumitem}
\usepackage{listings}
\usepackage{float}
\usepackage{mathtools}
\usepackage{tikz}
\usetikzlibrary{external}
\usepackage{trimclip}
\usepackage{afterpage}
\usetikzlibrary{patterns}
\usepackage{balance}
\usepackage{ragged2e}
\usetikzlibrary{plotmarks}

\newcommand*{\sNNcc}{\ensuremath{\sqrt{s_\mathrm{NN}}=\SI{200}{GeV}}\xspace}
\newcommand*{\fmc}[1]{\ensuremath{#1\,\mathrm{fm}/c}\xspace}
\newcommand*{\erho}{\ensuremath{\varepsilon}\xspace}
\newcommand*{\pT}{\ensuremath{p_\mathrm{T}}\xspace}
\newcommand*{\FNOper}{\ensuremath{\mathrm{FNO}_{40}^{60}}\xspace}
\newcommand*{\FNOcen}{\ensuremath{\mathrm{FNO}_{0}^{10}}\xspace}
\newcommand*{\FNOcenper}{\ensuremath{\mathrm{FNO}_{0,40}^{10,60}}\xspace}
\newcommand*{\FNOall}{\ensuremath{\mathrm{{}^{0.8,1.12nw}FNO}_{0,40}^{10,60}}\xspace}
\newcommand*{\ICper}{\ensuremath{\mathrm{IC}_{40}^{60}}\xspace}
\newcommand*{\ICcen}{\ensuremath{\mathrm{IC}_{0}^{10}}\xspace}
\newcommand*{\ICcenspikey}{\ensuremath{\prescript{\mathrm{0.8\,nw}}{}{\mathrm{IC}_{0}^{10}}}\xspace}
\newcommand*{\ICperspikey}{\ensuremath{\prescript{\mathrm{0.8\,nw}}{}{\mathrm{IC}_{40}^{60}}}\xspace}
\newcommand*{\ICmidspike}{\ensuremath{\prescript{\mathrm{0.96\,nw}}{}{\mathrm{IC}_{40}^{60}}}\xspace}
\newcommand*{\ICcenmidspike}{\ensuremath{\prescript{\mathrm{0.96\,nw}}{}{\mathrm{IC}_{0}^{10}}}\xspace}
\newcommand*{\JETSCAPE}{\texttt{JETSCAPE}\xspace}
\newcommand*{\iSpectraSampler}{\texttt{SpectraSampler}\xspace}
\newcommand*{\MUSIC}{\textsc{music}\xspace}
\newcommand*{\MUSICcen}{\ensuremath{\textsc{MUSIC}_{0}^{10}}\xspace}
\newcommand*{\MUSICper}{\ensuremath{\textsc{MUSIC}_{40}^{60}}\xspace}
\newcommand*{\MUSICcenspikey}{\ensuremath{{}^\mathrm{0.8nw}\textsc{MUSIC}_{0}^{10}}\xspace}
\newcommand*{\MUSICperspikey}{\ensuremath{{}^\mathrm{0.8nw}\textsc{MUSIC}_{40}^{60}}\xspace}

\newcommand*{\NeuralOperator}{\textsc{NeuralOperator}\xspace}
\newcommand*{\eg}{\textit{e.g.,}\xspace}
\newcommand*{\ie}{\textit{i.e.,}\xspace}

\newcommand*{\ttau}{\ensuremath{\tau}\xspace}
\newcommand*{\PyTorch}{\texttt{PyTorch}\xspace}
\newcommand*{\xy}{$x$-$y$\xspace}
\newcommand*{\vx}{\ensuremath{v_{x}}\xspace}
\newcommand*{\vy}{\ensuremath{v_{x}}\xspace}
\newcommand*{\antikT}{\ensuremath{\mathrm{anti\text{-}}k_\mathrm{T}}\xspace}
\newcommand*{\vtwo}{\ensuremath{\mathrm{v_{2}}\xspace}}
\newcommand*{\GeVc}[1]{\SI{#1}{GeV/\mathit{c}}\xspace}
\newcommand*{\GeVcube}[1]{\SI{#1}{GeV/fm^{3}}\xspace}

\newcommand*{\singlefigure}[4]{
\begin{figure}[htbp]
    \centering
    \resizebox{1.0\linewidth}{!}{\clipbox{0.3cm 0cm 1.7cm 0cm}{#1}}
    \begin{tikzpicture}[overlay, remember picture]
        #2
    \end{tikzpicture}
    \vspace{-2em}
    \caption{#3}
    \label{#4}
\end{figure}
}

\newcommand{\doublefigure}[9][htbp]{

    \begin{figure*}[#1]
        \begin{minipage}{1.00\linewidth}
            \begin{minipage}{0.48\linewidth}
                \resizebox{1.00\linewidth}{!}{\clipbox{0.3cm 0cm 1.7cm 0.0cm}{#2}}
                \begin{tikzpicture}[overlay, remember picture]
                    #3
                \end{tikzpicture}
            \end{minipage}
            \begin{minipage}{0.48\linewidth}
                \resizebox{1.0\linewidth}{!}{\clipbox{0.3cm 0cm 1.7cm 0cm}{#5}}
                \begin{tikzpicture}[overlay, remember picture]
                    #6
                \end{tikzpicture}
            \end{minipage}\\
            \raggedright
            \hspace{1.1cm}
            \begin{minipage}[t]{0.42\linewidth}
                \justifying
                \hspace{-3em}(a) #4
            \end{minipage}
            \hspace{1.0cm}
            \begin{minipage}[t]{0.42\linewidth}
                \justifying
                \hspace{-3.3em} (b) #7
            \end{minipage}
        \end{minipage}
        \caption{\justifying #8}
        \label{#9}
    \end{figure*}
}

\newcommand*{\LatexJetscape}{
    \node[black, font=\fontsize{7.8}{15}\selectfont, rotate=0, align=left] at (0.3,7.67) {$\rm JETSCAPE\;Au+Au\;\sqrt{\mathit{s}_{NN}}=200\;GeV$};
}
\newcommand*{\LatexEta} {
    \node[black, font=\fontsize{7.6}{15}\selectfont, rotate=0, align=left] at (-1.8,7.37) {$|\eta|<1$};
}

\newcommand*{\LabPhi}{
    \node[black, font=\fontsize{16.8}{16}\selectfont, rotate=90, align=left] at (-3.95,7.14) {$\rm \frac{1}{N_{event}} \frac{dN}{d\phi}$};
    \node[black, font=\fontsize{10.5}{15}\selectfont, rotate=0, align=left] at (3.70,0.85) {$\phi\;\rm [rad]$};
    \LatexJetscape
    \LatexEta
}
\newcommand*{\LabPhiRat}{
    \node[black, font=\fontsize{11.2}{16}\selectfont, rotate=90, align=left] at (-3.69,6.83) {$\rm FNO/MUSIC$};
    \node[black, font=\fontsize{10.5}{15}\selectfont, rotate=0, align=left] at (3.70,0.85) {$\phi\;\rm [rad]$};
    \LatexJetscape
    \LatexEta
}
\newcommand*{\LabBulkPt}{
    \node[black, font=\fontsize{16.8}{16}\selectfont, rotate=90, align=left] at (-3.95,7.14) {$\rm \frac{1}{N_{event}} \frac{dN}{d\mathit{p}_{T}}$};
    \node[black, font=\fontsize{10.5}{15}\selectfont, rotate=0, align=left] at (3.25,0.85) {$\rm \mathit{p}_{T}\;[GeV/\mathit{c}]$};
    \LatexJetscape
    \LatexEta
}
\newcommand*{\LabBulkPtRat}{
    \node[black, font=\fontsize{10.5}{15}\selectfont, rotate=0, align=left] at (3.25,0.85) {$\rm \mathit{p}_{T}\;[GeV/\mathit{c}]$};
    \node[black, font=\fontsize{11.2}{16}\selectfont, rotate=90, align=left] at (-3.69,6.83) {$\rm FNO/MUSIC$};
    \LatexJetscape
    \LatexEta
}
\newcommand*{\JetSeven}{
    \node[black, font=\fontsize{7.6}{15}\selectfont, rotate=0, align=left] at (-0.25,7.37) {$\rm anti\mbox{-}\mathit{k}_{T}\;\;R=0.7\;\;jets\;\;|\eta_{jet}|\le1$};
}
\newcommand*{\LabJetPt}{
    \node[black, font=\fontsize{16.8}{16}\selectfont, rotate=90, align=left] at (-3.95,7.34) {$\rm \frac{1}{N_{jet}} \frac{dN}{d\mathit{p}_{T}}$};
    \node[black, font=\fontsize{10.5}{15}\selectfont, rotate=0, align=left] at (3.25,0.85) {$\rm \mathit{p}_{T}^{jet}\;[GeV/\mathit{c}]$};
    \LatexJetscape
    \JetSeven
}
\newcommand*{\LabJetPtRat}{
    \node[black, font=\fontsize{10.5}{15}\selectfont, rotate=0, align=left] at (3.25,0.85) {$\rm \mathit{p}_{T}^{jet}\;[GeV/\mathit{c}]$};
    \node[black, font=\fontsize{11.2}{16}\selectfont, rotate=90, align=left] at (-3.69,6.83) {$\rm FNO/MUSIC$};
    \node[black, font=\fontsize{7.8}{15}\selectfont, rotate=0, align=left] at (0.8,7.67) {$\rm JETSCAPE\;Au+Au\;\sqrt{\mathit{s}_{NN}}=200\;GeV$};
    \node[black, font=\fontsize{7.6}{15}\selectfont, rotate=0, align=left] at ( 0.31,7.37) {$\rm anti\mbox{-}\mathit{k}_{T}\;\;R=0.7\;\;jets\;\;|\eta_{jet}|\le1$};
}

\newcommand*{\LabJetZ}{
    \node[black, font=\fontsize{16.8}{16}\selectfont, rotate=90, align=left] at (-3.95,7.34) {$\rm \frac{1}{N_{jet}} \frac{dN}{dz}$};
    \node[black, font=\fontsize{10.5}{15}\selectfont, rotate=0, align=left] at (4.05,0.85) {$\rm z$};
    \LatexJetscape
    \JetSeven
}
\newcommand*{\LabJetZrat}{
    \node[black, font=\fontsize{11.2}{16}\selectfont, rotate=90, align=left] at (-3.69,6.83) {$\rm FNO/MUSIC$};
    \node[black, font=\fontsize{10.5}{15}\selectfont, rotate=0, align=left] at (4.05,0.85) {$\rm z$};
    \LatexJetscape
    \JetSeven
}

\begin{document}
\preprint{APS/123-QED} 

\title{
Fast prediction of the hydrodynamic QGP evolution in ultra-relativistic heavy-ion collisions using  Fourier Neural Operators }

\author{D.~Stewart}\affiliation{Wayne State University, Detroit, Michigan 48201}
\author{J.~Putschke}\affiliation{Wayne State University, Detroit, Michigan 48201}

 

\date{\today}

\begin{abstract} 

\noindent 

Recent research in machine learning has employed neural networks to learn mappings between function spaces on bounded domains termed ``neural operators''. As such, these operators can provide alternatives to standard numerical methods for partial differential equation (PDE) solutions. In particular, the Fourier Neural Operator (FNO) has been shown to map solutions for classical fluid flow problems with accuracy competitive with traditional PDE solvers and with much greater computing speed. This paper explores the first application of FNOs to model ultra-relativistic hydrodynamic flow of the quark-gluon plasma (QGP) generated in relativistic heavy-ion collisions. The application in ultra-relativistic flow is novel relative to classical flow, due to the hydrodynamic evolution of the QGP occurring in femtometer-scaled explosions characterized by rapid expansion cooling.  In this study we investigate the applicability of FNOs as computationally fast alternatives to standard numerical PDE solvers. The FNO predictions are evaluated by comparing to standard PDE solutions, using \MUSIC in the \JETSCAPE Monte Carlo event generator framework. The performance of calculating established experimental observables for flow and jet quenching using FNOs in the MC framework are also reported.

\end{abstract}
\maketitle

%
%
%
%
%

\section{Introduction}\label{sec:Introduction}

\subsection{The Quark Gluon Plasma (QGP)}

Ultra-relativistic heavy-ion collisions were motivated as a means to create the densities and temperatures necessary to form the quark-gluon plasma (QGP) a novel phase of matter predicted by quantum chromo-dynamics (QCD) and understood to have been the dominant state of matter in the first microseconds following the Big Bang.  From its first observations at the Relativistic Heavy Ion Collider (RHIC) at the Brookhaven National Laboratory in 2001 \cite{BRAHMS:2004adc,PHOBOS:2004zne,STAR:2005gfr,PHENIX:2004vcz}, and then at higher energies at the Large Hadron Collider (LHC) at CERN in 2009 \cite{ATLAS:2010isq,CMS:2011iwn,ALICE:2010suc}, the study of the QGP has remained a focus of intense research. In addition to understanding the QGP itself, principle motivations include using the QGP to understand the QCD phase diagram, cosmological evolution's sensitivity to and interplay with QCD, and the formation processes of baryonic matter, \ie  ``hadronization'' \cite{Busza:2018rrf}. 

The first observed signatures of the QGP, flow and jet quenching, continue to comprise the foundation of experimental measurements of the QGP today. As this paper will use Monte Carlo (MC) models to simulate both, they are briefly described below.

Spatial anisotropy in a heavy-ion collision geometry result in pressure gradients in the QGP's initial conditions (IC). In response, flow in the evolving QGP converts the pressure gradients into momentum gradients whose signatures persist in the final-state particles formed by the QGP's hadronization. The observation of a clear second order Fourier coefficient ($v_{2}$) in the azimuthal distribution of hadrons in off-center (``peripheral'') Au+Au collisions was one of the first indications of QGP formation.

In high energy collisions rare high-momentum-transfer partonic scatterings result in collimated sprays of high transverse-momentum (\pT) particles which are algorithmically clustered into objects called jets. In heavy-ion collisions the high energy partons may traverse the evolving QGP, and consequently undergo color force interactions, scattering and induced gluon emission, in the medium. This is jet quenching: the modification of jets and their constituent particles in heavy-ion collisions, relative to jets in $pp$ collisions. The clear suppression of high-\pT hadrons, and later jets, in head-on (``central'') Au+Au collisions was the other early evidence of QGP formation.

From these initial, clear indications of QGP formation, flow and jet measurements have been leveraged with increasingly sophisticated experimental and modeling techniques to constrain QGP physics (see \cite{Niida:2021wut,KrizkovaGajdosova:2020pxc,Cunqueiro:2021wls}). These involve modeling the complex evolution of heavy-ion collisions. This evolution includes: the initial configuration of ions \cite{Eskola:2016oht}; orientation and impact parameter of the colliding ions \cite{Miller:2007ri}; energy deposition to form the QGP ICs \cite{Krasnitz:1999wc,Shen:2017bsr}; evolution and flow of the QGP through freeze-out (this paper); hadronization \cite{Fries:2025jfi}; and finally the expansion and cooling of the hadron gas into free-streaming particles \cite{Altmann:2024kwx}. Monte Carlo (MC) simulations are used to capture the complexities of the interacting processes and compare theory to experiment. \JETSCAPE, used in this paper's investigation, provides a framework for these MCs; it simulates an entire heavy ion collision using modular compositions of smaller, process-specific,  MCs \cite{Putschke:2019yrg}. 

To use a MC, such as those provided by \JETSCAPE, and constrain theory with measurement, one typically conducts a parametric study varying the physics parameter(s) under question and observing the effects on the measured particle distributions.  These methods tend to be statistics hungry, requiring significant numbers of events to cover the phase space of the studied parameters.  Other applications include global Bayesian analyses, in which multiple parameters within the individual MC processes are constrained simultaneously with heavy-ion collision measurements (\eg \cite{JETSCAPE:2020mzn}). Due to the high dimensional phase space covered in the simultaneous optimization, Bayesian analyses require even larger statistical samples.

In these MC calculations, QGP hydrodynamic flow is governed by coupled partial differential equations (PDEs) derived from Quantum Chromodynamics (QCD). Solving these PDEs with numerical methods is the most computationally expensive part of the simulations, typically by around an order of magnitude relative to all the other calculations. Various strategies are used to mitigate the constraint imposed by the resources required to calculate flow, such as ``oversampling'', in which the time step history of each hydrodynamic evolution is saved and reused for multiple events in which other parameters are individually varied; $e.g.$, the QGP evolution of each event is reused to simulate the quenching of multiple different jets. Regardless of the strategies employed, the computational cost of calculating QGP hydrodynamic flow is typically the limiting factor in the statistical power of any MC study in which flow is included. In this paper we explore a machine learning technique, Fourier Neural Operators (FNOs), as a computationally fast alternative to standard numerical PDE solvers.

\subsection{Relativistic Hydrodynamics}

The success of relativistic hydrodynamics in describing flow observables in experiment has made it the default tool to describe and experimentally constrain many-body QCD dynamics, making it the ``standard model'' for QGP flow in theory and modeling \cite{Shen:2020mgh}. In these measurements, the QGP has been observed to have a shear viscosity-to-entropy ratio approaching the theoretical minimum,  resulting in the most perfect fluid ever created in the laboratory \cite{Heinz:2005zg}.

The QGP flow takes place in a dynamic system, formed by two ions colliding with initial opposing velocities close to the speed of light ($c$) along the beam line (the $z$ axis in the geometry used in this study). Consequently, when the QGP is formed, it is already expanding along the beam line at approximately $c$. The QGP develops a transverse velocity from its immense internal energy density (\erho) and pressures, which also approach $c$ by the time of freeze-out.

This present study uses the relativistic hydrodynamic code \MUSIC \cite{Schenke:2010nt,Schenke:2010rr,Paquet:2015lta} to evaluate the  QGP flow in plane transverse to $z$ at the point of the collision (the ``\xy plane''). As is customary in studying relativistic hydrodynamics, \MUSIC uses Milne coordinates, in which the variables time $t$ and longitudinal coordinate $z$ are translated into the longitudinal proper time $\tau$ and the space-time rapidity $\eta_s$ as follows: 
\begin{align} 
\tau = \sqrt{t^2 - z^2}, \eta_s = \frac{1}{2} \ln \frac{t + z}{t - z}. 
\end{align} 
This formulation makes $\tau$ a numerically efficient way to capture the rapid expansion of the unit fluid cells along the $z$ axis, a fact that this paper's study will exploit. 

The hydrodynamic equations for energy-momentum conservation in this coordinate system take the following form,
\begin{align}
    \partial_\tau (\tau T^{\tau \tau}) + \partial_i (\tau T^{i \tau}) + \partial_{\eta_s} T^{\eta_s \tau} + T^{\eta_s \eta_s} &= 0 \label{eq:EOM1} \\
    \partial_\tau (\tau T^{\tau j}) + \partial_i (\tau T^{i j}) + \partial_{\eta_s} T^{\eta_s j} &= 0  \label{eq:EOM2} \\
    \partial_\tau (\tau T^{\tau \eta_s}) + \partial_i (\tau T^{i \eta_s}) + \partial_{\eta_s} T^{\eta_s \eta_s} + T^{\tau \eta_s} &= 0, \label{eq:EOM3}
\end{align}
where $T^{\mu\nu}(\tau, x, y, \eta_s)$ is the system's energy-momentum tensor and indices $i, j$ run over the $x$ and $y$ directions in the transverse plane. The PDEs in Eqs.~\eqref{eq:EOM1} to \eqref{eq:EOM3} differ from the conventional hydrodynamic equations which are expressed in Cartesian coordinates; \eg the coordinate transformation results in an extra extra $T^{\eta_s \eta_s}$ term in Eq.~\eqref{eq:EOM1}. \MUSIC evolves the QGP by numerically solving these coupled PDEs.

\subsection{Fourier Neural Operators}

Within the past few years, machine learning techniques called neural operators have been developed to learn mappings between function spaces on bounded domains \cite{JMLR:v24:21-1524}. A novel feature of these techniques is that because they are mapping functions, they may be trained with data (function parameters) using any set of discretizations and then used to predict data at any other independently-selected set of discretizations. This differs from conventional convolutional neural networks, which learn mappings between fixed-size input and outputs. 

Neural operators have promising potential as alternatives to numerical PDE solvers in physics and engineering applications, where solutions to known PDEs are often intractable and may also be prohibitively expensive to measure via physical experiments \cite{Azizzadenesheli2024}. A Python library called \NeuralOperator, built upon \PyTorch, to efficiently train neural operators using an accessible interface, is used in this study \cite{kossaifi2025librarylearningneuraloperators}).

One published neural operator implementation, the Fourier Neural Operator (FNO), lifts the parameters into Fourier space before parameterizing for the neural operator mapping \cite{li2021fourierneuraloperatorparametric,JMLR:v24:21-1524}. Studies have demonstrated that FNOs are a powerful tool for approximating the solution to the PDEs of classical fluid flow problems governed by the Navier-Stokes equations, a 1D viscous Burgers equation, and Darcy's flow \cite{Li2021Fourier}. In these studies, FNOs are trained using hydrodynamic flow calculated by PDE solvers for a variety of fluid ICs, and then tested against PDE solutions for other ICs, and are demonstrated to be effective proxies for the PDE solvers \cite{JMLR:v24:21-1524}. This includes ``spatial super-resolution'' (or discretization-invariance), in which an FNO is trained on a lower spatial resolution, and then tested at a higher spatial resolution. 

For an additional, interesting published FNO application the reader is also referred to \cite{pathak2022fourcastnetglobaldatadrivenhighresolution}, in which FNOs' discretization-invariance is used with great effect on multi-scale and inhomogeneous global weather input data to produce high resolution global weather predictions. With this referral, we also note that there are parallels to the multi-scale and inhomogeneous nature of heavy-ion collision detectors and data.

Motivated by this success of FNOs in modeling classical hydrodynamic flow, this paper presents the first investigation into using FNOs to model ultra-relativistic hydrodynamic flow of QGP in heavy-ion collisions.  If applicable, FNOs would provide a fast, accurate, discretization-invariant alternative for PDE solvers to calculate the hydrodynamic evolution of the QGP in MC simulations. 

\subsection{Advantages of FNOs for QGP Flow}

If validated, FNOs can provide an alternative for standard numerical solutions for QGP flow in relativistic heavy-ion collisions. The most crucial advantage would be a dramatic decrease in computational time and resources with respect to current MC implementations. This would allow higher statistical samples to be produced, as well as enabling ``exploratory'' calculations which could be performed on personal computers/laptops rather than computing clusters, speeding up the development of new theoretical approaches and the exploration of new experimental observables using realistic MC event generators.

The value of computationally fast implementations is crucial in applications requiring PDEs which are more expensive to solve. For example, many flow simulations, as is the case with those used in this current study, are ``2+1D'': \ie they are solved in the 2-dimensional plane transverse to the collision geometry and evolved along the time dimension. This approximation is good at ``mid-rapidity'', but becomes less accurate with increasing rapidity representing the longitudinal expansion of the QGP along the beam axis. Such predictions would require a true 3+1D geometry, and would be much more computationally expensive to simulate, and thereby increasing the value of a fast and accurate alternative. The need for such simulation will likely only increase with the increasing interest, and evolving experimental detectors, to measure rapidity-differentiated physics and forward physics (\eg \cite{LHCForwardPhysicsWorkingGroup:2016ote,Hentschinski:2022xnd}). 

FNOs are also discretization-invariant; i.e., they may be queried at any spatial precision. This may prove particularly convenient for calculating higher flow harmonics, such as $v_3$, $v_4$ ..., $v_n$, while reducing the amount of computing time for the training samples since they can be produced at a lower spatial resolution.

Additionally, the FNOs predict the entire hydrodynamic evolution history. This is necessary for any realistic jet quenching MC implementation. FNOs have also been shown, in their original work, to be able to recursively predict future time steps in the hydrodynamic evolution. Such a feature, if demonstrated in the context of the QGP, would allow a full concurrent evolution of the QGP medium including energy deposition/medium feedback of jets due to partonic energy loss. In such a scenario common approaches to reuse the hydro evolution to reduce computation time is not applicable and fast and efficient alternatives like FNOs might become necessary.

\subsection{Study Introduction and Overview}

This study is intended as an initial investigation into using FNO as  alternatives for standard numerical solvers for QGP flow in heavy-ion collisions. Therefore, existing software, tools, and libraries are used whenever possible. The custom code and calculation workbooks are deliberately minimal and straight-forward to be easily understandable. These are archived at \cite{githubarch} with the intention that they may be used as resources in future more extensive studies.

The goal of this investigation is to demonstrate if FNOs are practicable in this application and, if they are, to illustrate their performance. To do so, this study uses \JETSCAPE to generate MC simulations of Au+Au collisions at \sNNcc. The ``truth-level'' QGP flow is calculated by \MUSIC, and sets of events using different collision parameters are used to train different FNOs. The FNOs' flow predictions are evaluated by direct comparison to \MUSIC's calculations. To then investigate performance in physics' calculations, \JETSCAPE freezes-out evolved QGPs (multiple times for each event, once using \MUSIC, and again using various FNOs) into final-state particles (using the \iSpectraSampler library \cite{Shen:2014vra,Denicol:2018wdp}) which are collectively called ``hadrons'' in this report.

The hadrons produced are used to study flow by inspecting their azimuthal distribution  and transverse momentum (\pT) distribution at mid-$\eta$. These experimental ``flow'' variables result from the QGP's hydrodynamic flow resulting from anisotropies in peripheral events. The experimental flow measured using the FNOs is compared to the flow using \MUSIC.

FNOs effects on physics measurements are also studied in the context of partonic energy loss in the QGP. The high energy partonic scatterings that result in jets are formed early in heavy-ion collisions. This study embeds \GeVc{20} gluons, with momentum in random azimuths, into the \xy plane center of central collisions. As is the case in nature, the quenching experienced by the gluons in the MC occurs throughout the entire evolution of the QGP (for this reason jets serve as a kind of tomographic probe of the QGP \cite{Apolinario:2022vzg}). The hadronized particles resulting from the embedded gluons are clustered into jets. The quenching measured when using FNOs in the MC is compared to that that resulting from using \MUSIC.

These measurements are presented in detail in the remainder of this paper, which is organized as follows. The final section of this introduction (Sec.~\ref{sec:caveats}) discusses simplifications used in this study. Section~\ref{sec:implementation} presents and discusses the implementation details and algorithms used in the FNO training. The methodology and results for evaluating the FNOs are presented in Sec.~\ref{sec:evaluation}, while the methodology and results investigating FNO performance in simulated flow and jet quenching measurements are presented in Sec.~\ref{sec:verification}. Finally, Sec.~\ref{sec:conclusions} gives the conclusions and outlook.

\subsection{Study Simplifications and (current) Limitations}\label{sec:caveats}

In this study several limitations, and consequently simplifications, are introduced.  These are mainly due to limited computational resources available (see Sec.~\ref{app:hardware}) with the most limiting being the rather modest GPU memory. They are addressed in the following sections and also listed here for convenience:

\begin{itemize}
\item ``Ideal'', i.e. non-viscous, hydrodynamic flow is used. This reduces the required number of QGP modeling parameters on the 2D+1 grid from nine to three, making the model size significantly more tractable for training and implementation. 

The effect is a small modification of the hadronization process of the QGP at freeze out. The effects are anticipated to be small on the simulated flow measurement (modification of hadronic \pT and azimuthal distributions), and even smaller on jet quenching. Regardless, the importance to this paper is not the flow or quenching, \textit{per se}, but the relative sensitivity of the calculation to using the FNO as a proxy for \MUSIC.

\item In the quenching MC, gluon jets are embedded, even though at these kinematics quark initiated jets are much more common. Gluons were chosen because they experience stronger quenching effects, and therefore will be more sensitive to differences in QGP flow introduced by the FNOs. As such, the simulated quenching measurement is not itself representative of experimental results but is instead a more sensitive indicator of FNO performance.

\item The FNOs are only trained for the first \fmc{6} following the QGP formation. This is a result only of the memory limitation of extending the FNO training (and implementation) to additional steps in \ttau. Fortunately, this is sufficient to simulate more than 97\% of peripheral events up to the point of hadronization, as is required for the flow measurements. 

The limitation to \fmc{6} also introduces an event selection bias when simulating peripheral events with smaller nucleon widths. In those events, more events are discarded because they do not reach freeze-out. However, the necessary comparisons are not between events using different nucleon widths, but rather between the results of FNOs and \MUSIC for the same events at each individual nucleon width.

Jet quenching is studied in central events using only these first \fmc{6} of QGP flow, when most of the jet quenching occurs. However, to allow a proper comparison we also use only the first \fmc{6} of \MUSIC.

\item The ``super-resolution'' of a FNO predicting results on a $150\times150$ \xy grid (even though it is trained only on $60\times60$ \xy grids) is evaluated directly against the \MUSIC predictions, but only out to \fmc{5}. It is not hadronized to compare in simulated physics measurements. Again, this was motivated simply by the limitations of the size and complexity introduced by the more than six-fold increase in grid-size.

\item The physics simulation measurements use the hadrons directly as provided by the MC without any further particle identification or transverse momentum cuts. Neither detector acceptance effects or uncertainties are included. As such, all errors presented are simple statistical uncertainties.

\end{itemize}

\section{Methodology: MC Simulation and NFO Training}\label{sec:implementation}

This section presents and discusses the methodologies for generating data with the MCs, and training the NFOs.

\subsection{Implementation Relative to Classical Hydrodynamics}\label{subsec:compclass}

This study trains FNOs using the pattern published using NFOs for classical hydrodynamic flow. In one aspect, this is a  natural adaptation - after all, the QGP flows as a ``perfect fluid''.  On the other hand, the relativistic flow is very unique. The flow in each event occurs in an explosion at exceptional temperatures and pressures, a ``little-bang'' so to speak \cite{Sun:2022xjr}, proceeding from the QGP's IC from the collision and lasting on the order of \fmc{10}.

Both this similarity and uniqueness to classical hydrodynamics are reflected in the implementation of the FNOs in this study. As was done in the studies of classical hydrodynamics, this study takes collections of ICs, propagates flow using a PDE (in this case \MUSIC), and trains on the saved flow evolution data.

Unlike the hydro studies for classical flow, the ICs for QGP flow are also determined by the parameters used in the ion collisions which produce them. If the FNO really were a perfect proxy for the PDE solver (\MUSIC), then it would be able to predict the flow for any IC resulting from any set of parameters used in the MC. However, PDE solutions (and therefore proxies) can be highly sensitive to the phase space of their ICs.

In this study, we investigate two MC parameters on which the IC phase space is sensitive.  The first is the centrality of the collisions.  Centrality is defined by percentiles of the impact parameter between the centers of the colliding nuclei. We use two ranges of centrality: central events (0-10\%), and peripheral events (40-60\%), in which the impact parameter percentiles are calculated using unbiased collisions from colliding ion gases.

Central collisions result in the highest multiplicities. They also start with the largest QGP volume and highest \erho values, and evolve for the longest time prior to freeze-out. Consequently, they also exhibit the most jet quenching. The peripheral collisions correspondingly have lower (smaller) multiplicities, \erho values, QGP volumes, and evolution times. However, they have higher physical anisotropies, resulting in larger values of flow (\vtwo).

Our computing resources were large enough to make FNOs which map \fmc{6} units from the QGP IC, which is sufficient to reach freeze-out for more than 97.5\% of peripheral events using the nominal nucleon width (the remaining events are discarded in this study). Fortunately, while measured flow is sensitive to the final state of the QGP prior to freeze-out, jet quenching is most sensitive to the high-\erho evolution of the QGP at smaller \ttau values. Consequently, in this study, we measure the jet quenching occurring only in the first \fmc{6.5} of central events, and flow only in peripheral events.

The second influential MC parameter for the QGP ICs is the nucleon width used in modeling the colliding ions. The proper nucleon width to use is a topic of ongoing research \cite{Nijs:2022rme} and therefore in MCs used in experiment it is common to use more that one value to help quantify and propagate the uncertainty from the nucleon width on the physics measurement. Smaller nucleon widths result in sharper, or ``spikier'', \erho distributions in QGP ICs.  This study uses a nominal value of \SI{1.12}{fm} but also investigates the results of using the smaller values of \SI{0.8}{fm} and \SI{0.96}{fm}.

The second difference between this study and using FNOs with classical hydrodynamics come from the effects of the rapidly expanding QGP, which leads to a strong time dependence of  \erho as well as the fluid cell velocities. In the classical studies, there is no particular time preference for the fluid parameters; they are approximately conserved on a system level between time steps. In the QGP, all flows start at relatively high-\erho and ends at freeze-out; with the predictions for quenching being more important at small to medium \ttau and flow at the final \ttau. Meanwhile \vx and \vy always start at zero, since for this study no pre-equilibrium dynamics \JETSCAPE module was used, and approach $|\vec{v}_{xy}|\approx{}c$ at the QGP boundaries.

As a result, classical hydrodynamic studies can use standard ML methods to normalize the data for training and applying the FNOs. In the course of this study two such normalization methods were adapted for the QGP little-bangs, in which all \ttau steps were independently normalized. In the end, it was determined that a simple normalization in which the \erho values at each \ttau step are scaled to account for the cooling from  longitudinal ($z$ axis) expansion has the best performance. This is discussed in more detail in Sec.~\ref{subsec:normalization}.

\subsection{Code and Libraries}

To enable convenient reproduction of this study and facilitate extension in potential future studies, existing published code and libraries are used for the MC simulations (\JETSCAPE), the PDE solver (the \MUSIC module in \JETSCAPE), training the FNO (the \NeuralOperator library \cite{kossaifi2025librarylearningneuraloperators}), and hadronizing the QGP (\iSpectraSampler, also via \JETSCAPE). All new code generated to conduct this study is archived and documented at \cite{githubarch}. The hardware used is listed in Sec.~\ref{app:hardware}.

\subsection{MC Data and FNO Training}\label{subsec:fno_training}

The general details of the FNO training are given here. For more detail refer to the archived code \cite{githubarch} and Sec.~\ref{app:training_details} in the appendix.

This study uses \JETSCAPE to generate peripheral and central \sNNcc Au+Au collisions using both nominal and smaller nucleon widths. For convenience, ICs produced from central and peripheral events are indicated by the centrality range, \eg \ICcen and \ICper, respectively. Events using smaller nucleon widths than the nominal value (\SI{1.2}{fm}) are indicated with a left side superscript, \eg \ICcenspikey for a \SI{0.8}{fm} nucleon width. Analogous indications are used to differentiate results of the MC using various FNO models (according to the ICs on which they were trained) and the MCs using \MUSIC, \eg \MUSICcen.

The IC of the QGP for each collision  are saved to disk. Additionally, \MUSIC evolves a non-viscous  QGP from the IC at $\tau=\fmc{0.5}$ throughout freeze-out, and 3 QGP properties (\erho, \vx, and \vy) are saved at intervals of \fmc{0.1}, the standard timestep size used in \JETSCAPE jet quenching modules, up to $\tau=\fmc{6.4}$ (59 steps from the IC). The data is saved on \xy grids of dimension \SI{30}{fm} square, with $60\times60$ measurement grid-points (\ie resolution of \SI{0.5}{fm}). As a result, the total memory for the fluid evolution for each event is a tensor of size $\left(n_x\times n_y\times n_{\tau}\times n_\mathrm{\erho,vx,vy}=60\times60\times60\times3=648,000\right)$, of which $1/60^\mathrm{th}$ is the IC. 

FNOs were trained with the following datasets:

\begin{itemize}
\item \FNOcen from 10,000 \ICcen events
\item \FNOper from 10,000 \ICper events
\item \FNOcenper from 5,000 \ICcen and 5,000 \ICper events
\item \FNOall from 2,000 events: 500 each of \ICcen, \ICper, \ICcenspikey, and \ICperspikey
\end{itemize}

Each FNO was trained using the \NeuralOperator library with the following options:
\begin{itemize}
    \item All data is normalized by scaling each \erho value by the associated \ttau value
    \item Use 80\% of the events for training,  the Sobolev norm for the training loss (refer to \cite{JMLR:v24:21-1524} and references therein)
    \item Use the remaining 20\% of events for validation, using the mean squared error
    (\texttt{L2Loss}) for the validation loss
    \item Use the AdamW optimizer \cite{loshchilov2019decoupledweightdecayregularization}
    \item Train with 250 epochs
    \item Define the FNO using the weights from the epoch with the minimum
    \texttt{L2Loss}
\end{itemize}

\subsection{Selection of FNO Normalization}\label{subsec:normalization}

As motivated in Sec.~\ref{subsec:compclass}, two standard ML normalizations were adapted for the QGP flow tensors prior to training the FNOs. Using $i$, $j$, and $k$, to index $x$, $y$, and $\tau$, respectively, these are:
\begin{itemize}
\item Gaussian normalization: normalize \erho in each time step so that the average ($\mu$) is zero with a standard deviation ($\sigma$) of unity. The normalization from $\erho$ to $\erho'$ is: $$\erho'_{i,j,k} = \frac{\erho_{i,j,k}-\mu_{k}}{\sigma_k}$$ This is a standard ML normalization, and indeed the one used for FNO training in classical hydrodynamics \cite{Li2021Fourier}.
\item Normalize the maximum, also a common ML normalization: $$\erho'_{i,j,k} = \frac{\erho_{i,j,k}}{\max\left(\erho_k\right)}$$
\end{itemize}

An additional two normalizations motivated by the physics were tested:
\begin{itemize}
\item Approximate constant entropy:  $$\erho_{i,j,k}'=\erho_{i,j,k}^{0.75}$$ this is motivated by the non-viscous hydro having constant entropy, which is approximately $\erho^{0.75}$. as such, it helps put all \ttau steps on the same footing.
\item Remove the cooling from axial expansion (i.e. expansion at approximately the speed of light along the $z$ axis). Because \ttau scales in units of $c$, this normalization is simply:  $$\erho'_{i,j,k}=\tau_{k}\times\erho_{i,j,k}$$
\end{itemize}

While other physics normalizations are possible and may be interesting for future study (\eg combining both physics inspired normalizations), it was observed that with sufficient statistics, training with any of above normalizations produced FNOs that appeared to capture \erho, \vx, and \vy as long as the FNO centrality matched the IC centrality. In general it seemed that the physics inspired normalizations better capture the shape of the \erho distributions when applied on data with MC parameters not included in the training samples. For an example, refer to Fig.~\ref{fig:cen_on_per} in the Appendix.

\section{FNO Evaluation}\label{sec:evaluation}

\subsection{Methodology}

Each FNO was tested against the \MUSIC evolution in independent MC events, using MCs for each kind of IC:  \ICcen, \ICper, and \ICperspikey. To do this, the mean values of the QGP velocity (both radially outward from, and curling around, the center of the \xy grid) were calculated for an array of radii and of values of \ttau. The results of the FNOs acting on the the ICs, ``FNO(IC)'', were compared against \MUSIC.

To verify the FNOs for the evolution of \erho, the values of \erho throughout the \xy grid for specific \ttau steps were inspected and compared to those generated with \MUSIC. To do this, constant valued \erho lines are drawn on the \xy grid (analogous to lines of constant elevation on a topographical map). The \erho values of the lines containing percentiles of the total event energies are calculated, using the $10^\mathrm{th}$, $20^\mathrm{th}$, $30^\mathrm{th}$, $40^\mathrm{th}$, and $60^\mathrm{th}$ percentiles.

The same process was also used to test the spatial super-resolution, in which an FNO trained on the $60\times60$ grid was then used to predict the flow using a $150\times150$ grid. The results were compared to the \MUSIC results using the same higher-resolution grid.

\subsection{FNO Predictions of \erho at Specific $\tau$}

\begin{figure*}
    \begin{minipage}{0.48\linewidth}
        \includegraphics[width=0.99\linewidth, trim=1cm 0cm 27cm 1cm, clip]{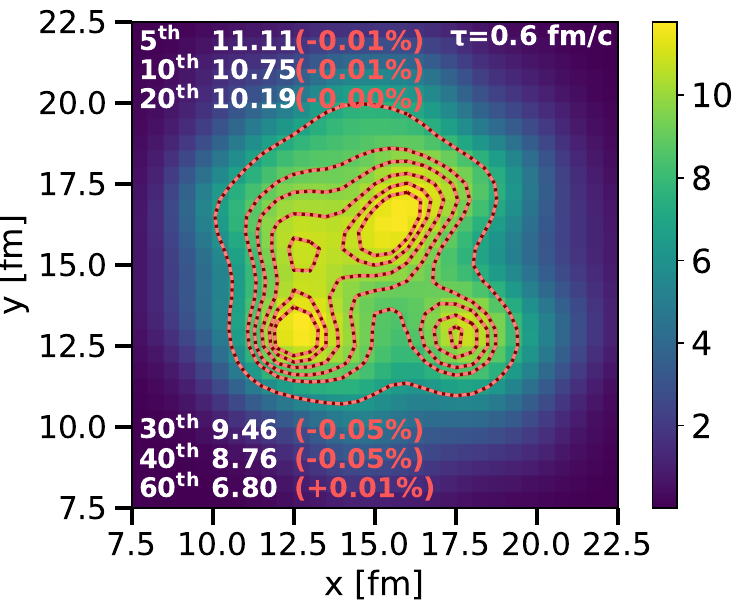}\\
         (a)  $\FNOcen(\ICcen)$, $\tau=\fmc{0.6}$
    \end{minipage}
    \begin{minipage}{0.48\linewidth}
        \includegraphics[width=0.99\linewidth, trim=26.7cm 0cm 1cm 1cm, clip]{central_0.pdf} \\
         (b)  $\FNOcen(\ICcen)$, $\tau=\fmc{6.4}$
    \end{minipage} \\
    \caption{
    \justifying Energy density (\erho) distributions of the QGP in a central \sNNcc MC Au+Au collision at first and last values $\tau$ predicted by the FNO. The
    color scale indicates the \erho distribution in
    $\mathrm{GeV}/\mathrm{fm}^3$ calculated by the PDE solver \MUSIC. The dotted lines plot constant \erho
    boundaries (analogous to elevation lines in a topographical map) containing percentiles of the total event energy.
    Values of \erho traced by the dotted lines are printed in white. 
    The location of the percentiles predicted by an FNO are shown in red lines, whose
    \erho values (relative to the \MUSIC values) are printed in red.}
    \label{fig:central_2D}
\end{figure*}

\begin{figure*}[htb!]
    \begin{minipage}{0.48\linewidth}
        \includegraphics[width=0.99\linewidth, trim=1cm 0cm 27cm 1cm, clip]{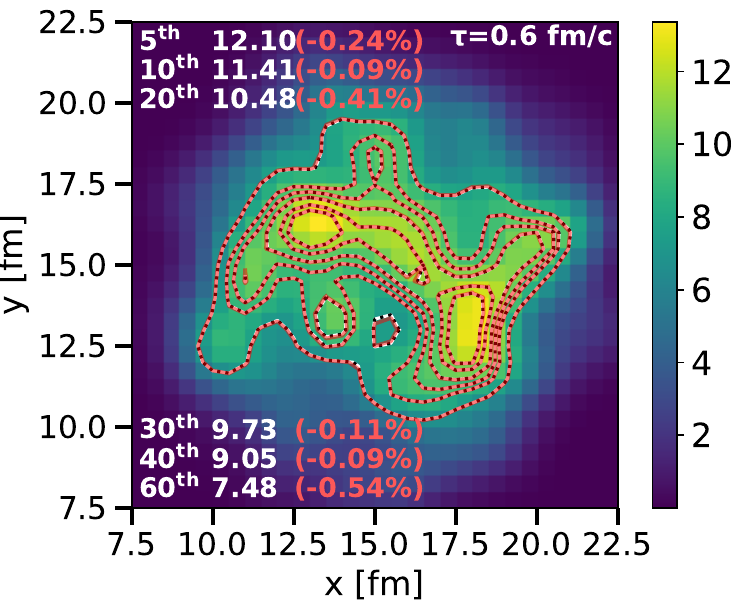}\\
         (a)  $\FNOcen(\ICcenspikey)$, $\tau=\fmc{0.6}$
    \end{minipage}
    \begin{minipage}{0.48\linewidth}
        \includegraphics[width=0.99\linewidth, trim=26.7cm 0cm 1cm 1cm, clip]{spiky_central_0.pdf}\\
         (b)  $\FNOcen(\ICcenspikey)$, $\tau=\fmc{6.4}$
    \end{minipage}
    \caption{
    \justifying Energy density (\erho) distributions of the QGP in a central \sNNcc MC Au+Au collision, with a nucleon width of \SI{0.8}{fm}, at the first and last values of \ttau predicted by the FNO. The color scale indicates the \erho distribution in $\mathrm{GeV}/\mathrm{fm}^3$ calculated by the PDE solver \MUSIC. The dotted lines plot constant \erho boundaries (analogous to elevation lines in a topographical map) containing percentiles of the total event energy. Values of \erho traced by the dotted lines are printed in white.  The location of the percentiles predicted by an FNO are shown in red lines, whose \erho values (relative to the \MUSIC values) are printed in red.}
    \label{fig:spiky_2D_central}
\end{figure*}

Fig.~\ref{fig:central_2D} shows the results for the first and final \ttau steps from $\FNOcen(\ICcen)$. The figure shows the 2D distribution of \erho, as well as the constant \erho contours containing the indicated percentiles of the total event energy. The \erho values and locations of the contours are shown for both the \MUSIC and FNO predictions. The results are typical of all predictions for which FNOs are applied on ICs using the same MC parameters (centrality and nucleon width) as were used to train the FNO. Namely, the \erho percentile value errors are typically sub-percentile, although they increase approximately monotonically with \ttau, and their contour shapes are very close to the truth contours. A figure using the same format as Fig.~\ref{fig:central_2D} is given in the Appendix in Fig.~\ref{fig:periph_2D} for $\FNOper(\ICper)$.

The results for an event with $\FNOcen(\ICcenspikey)$ are shown in Fig.~\ref{fig:spiky_2D_central}. In the figure, the generally sharper, spikier, initial distribution of \erho is apparent, along with the smoothing of the distribution by the final \ttau step. The shown results are also typical: the \FNOcen predictions are still quite good, but now contain errors in topography that are apparent to the eye, and errors in predicted values that are no-longer sub-percentile. Results for a typical peripheral event, using a smaller nucleon-width, is given in Fig.~\ref{fig:periph_2D_spikey} in the appendix using an FNO trained on central events with nominal nucleon widths ($\FNOcen(\ICperspikey$), which is essentially the maximum mismatch studied between ICs used in the NFO training and in the FNO application.

An FNO, \FNOall,  which is trained on only 500 events of each of the ICs (central and peripheral combined with nucleon width \SI{1.12} or \SI{0.96}) is applied to \ICcenspikey and found to be have quite good results. The errors in the \erho percentile values are generally less than 1\% in magnitude, and the visual errors on the plasma evolution smaller than those trained with the 10,000 \ICcen events. The results of the evolution of one event at \fmc{0.6}, \fmc{3.9}, and \fmc{6.4} are given in Fig.~\ref{fig:flow_FNOall_censpikey} in the appendix. 

When testing the same \FNOall on events with an intermediate nucleon width (\SI{0.96}{fm}), which is halfway-between the nucleon widths of the IC used to train \FNOall, the results are equally good as those for ICs from matching nucleon widths. A representative plot is given in Fig.~\ref{fig:erho_halfway}.

\subsection{FNO Evaluation for Super Resolution}

\begin{figure}[htbp]
    \centering
    \includegraphics[width=0.9\linewidth, trim=26.7cm 0cm 1.1cm 1cm,clip]{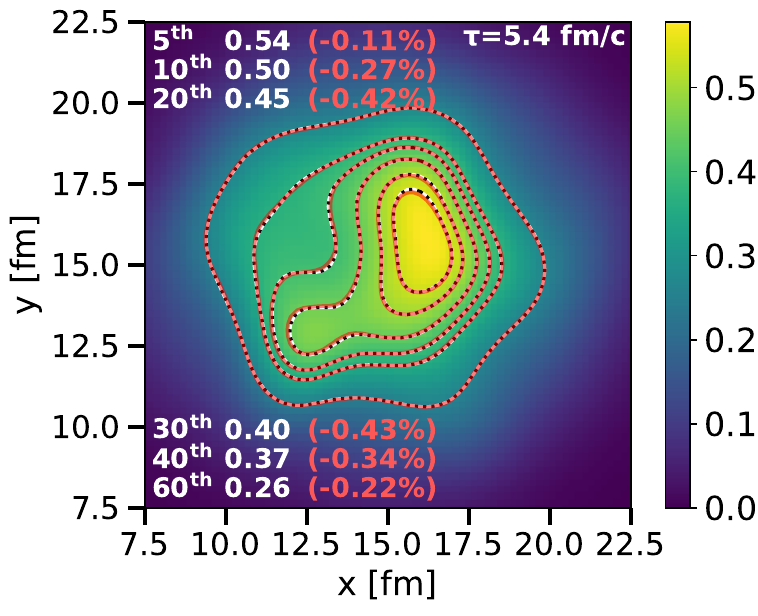}\\
    $\FNOcen(\ICcen)$, $\tau=\fmc{5.4}$
    \caption{Super-resolution: The results of using an FNO trained on central events with $60\times60$ resolution predicting values on a separate MC event with a resolution of $150\times150$. For an explanation of the legends and lines see caption of Fig.~\ref{fig:central_2D}.}
    \label{fig:superres}
\end{figure}

A separate FNO trained on central events, using 50 steps in \ttau (out to $\tau=\fmc{5.4}$), was used to make predictions against events using an $150\times150$ grid in \xy the IC. The performance of the hydrodynamic prediction is indistinguishable from the performance on the lower resolution ($60\times60$) grids on which it was were trained. The results at the final time step are shown in Fig.~\ref{fig:superres}. The closeness of the FNO predictions to those of \MUSIC show promise for generating hadron distributions for more sophisticated models (such as viscous-flow) at higher resolutions.

\subsection{FNO Predictions of $v_x$ and $v_y$}

\begin{figure}[htbp]
    \centering
    \includegraphics[width=0.9\linewidth, trim=0cm 0cm 0cm 0cm]{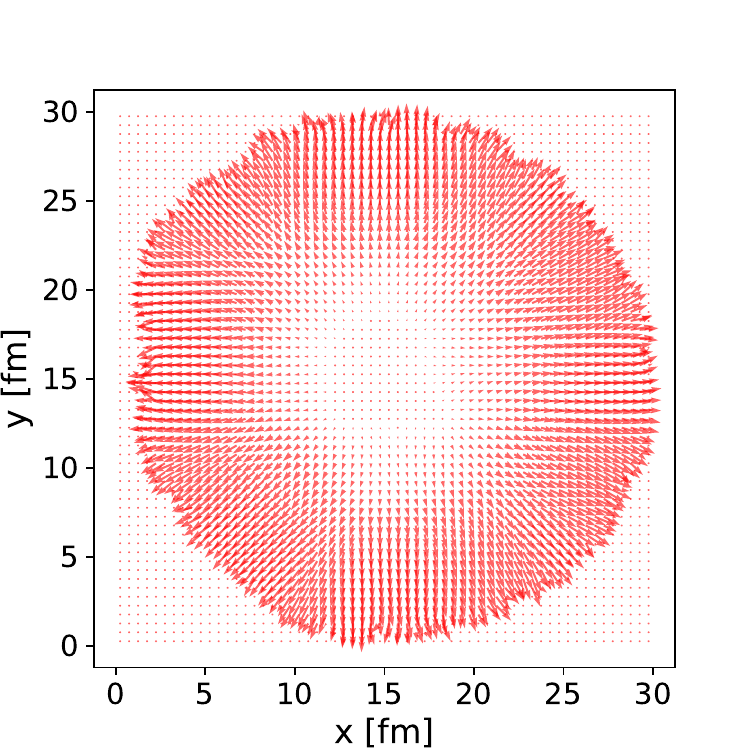}
    \caption{A vector field plot of the QGP velocities at $\tau=3.5$ in a
    central event. The values of the maximum length arrows are close to $c$.}
    \label{fig:vxvyvect}
\end{figure}

In each hydro evolution, the QGP starts with no velocity in the $x$-$y$ plane, and then rapidly expands outward, reaching velocities close to the $c$. The fastest velocities are at the surface of the expanding QGP. For example, the velocity vector field for a central event at $\tau=\fmc{3.5}$ is shown in Fig.~\ref{fig:vxvyvect}.

The values of the average distribution of radial velocity (relative to the center of the \xy center of the MC simulation) of two different annuli within $\tau\in[4.2,4.8]\;\mathrm{fm}/c$ are plotted in Fig.~\ref{fig:avg_vx_vy}. A figure showing the radial velocities for an array of annuli and \ttau is given in the appendix in Fig.~\ref{fig:dot_vel}. The distributions for the velocities curling around the center of the collision for the same radii and \ttau ranges are given in Fig.~\ref{fig:cross_vel}. In each instance, the mean velocities predicted by the FNO visually indistinguishable from the values calculated by \MUSIC.

\begin{figure}[htbp]
    \vspace{1em}
    \centering
        \includegraphics[width=0.7\linewidth, trim=2cm 4cm 15cm 3cm]{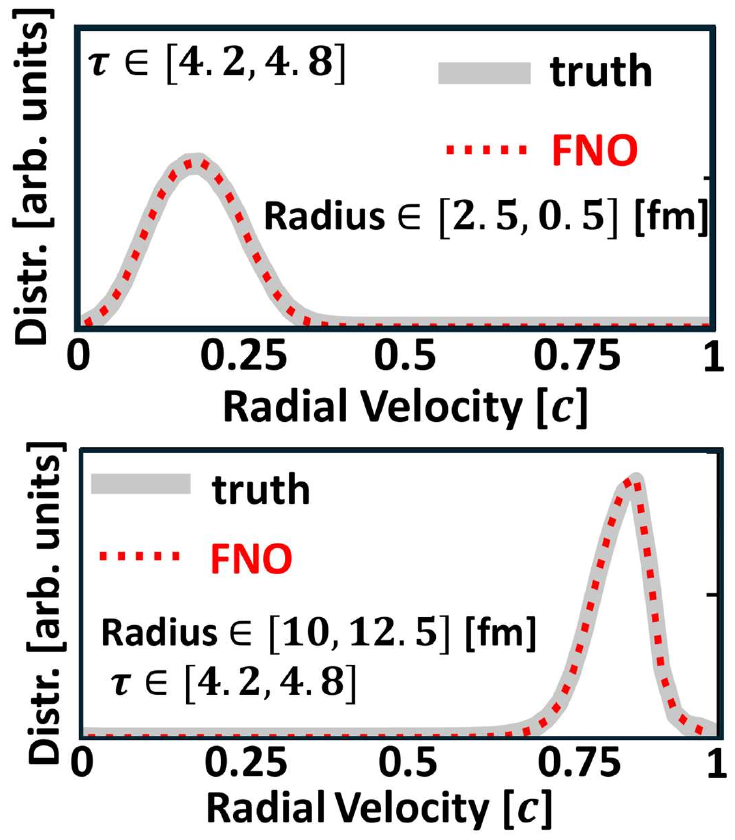}
        \caption{Average values of radial QGP velocity (relative to the $x$, $y$,
        center of collision in the MC) for 4000 central events at $x$ and $y$ grid points
        within the radii indicated and during the time evolution in $\tau$
        indicated.}
        \label{fig:avg_vx_vy}
\end{figure}

\section{Physics Performance Using FNOs}\label{sec:verification}

\subsection{Methodology}

Because of the complex interaction of physics processes in MC modules, an FNO's performance must be evaluated for specific desired physic measurements, using the final-state particles. This study evaluates performance of the FNOs when measuring azimuthal flow and jet quenching for FNOs for MC producing the different ICs.

To do this, new MC events were generated for the following ICs:
\begin{itemize}
\item 4,000 events with \ICcen
\item 4,000 events with \ICper
\item 2,000 events with \ICcenspikey
\item 2,000 events with \ICperspikey
\item 2,000 events with \ICmidspike
\end{itemize}

To evaluate the FNOs for measuring flow, \JETSCAPE used \iSpectraSampler to hadronize the QGP for all classes of peripheral events (\ICper, \ICmidspike, \ICperspikey) using the FNOs and separately using \MUSIC. All particles at mid-rapidity ($|\eta|<1$), without any \pT cut, were analyzed for their azimuthal and \pT distributions. To quantify the error relative to the truth values (i.e. \MUSIC), the ratios of the the FNO distributions to the \MUSIC distributions are also presented. In order to test the effects of FNOs to extrapolate to ICs between those used in the training, \FNOall was also tested on each of the peripheral ICs (one at each nucleon width) and compared to \MUSIC.

In order to evaluate the FNO performance for jet quenching, the central events were used, again with FNOs propagating various central ICs. Using the same ICs, \MUSIC generated the truth level data. Every event was used 10 times. Each time \JETSCAPE embedded a \GeVc{20} gluon into the IC, starting in the center of the \xy plane with a momentum trajectory at a randomly selected azimuthal angle. As noted above, in \sNNcc events, most \GeVc{20} partons are quarks, but gluons are selected for the embedding because they experience more quenching.

\JETSCAPE propagated the embedded gluons through the evolving QGP for \fmc{6} and hadronized the final state particles of only the jet. These hadrons are clustered with the \antikT algorithm using a jet resolution parameter $R=0.7$ \cite{Cacciari:2008gp}. Only the clustered jet with the highest-\pT in each event is considered. As a comparison baseline to show the quenching effects of the QGP, an additional set of events was simulated in which the jet interaction region was set to $0$, which a previous study showed are equivalent to \JETSCAPE jets generated using $pp$ collisions \cite{Stewart:2024mkx}.

Two standard observables to characterize the effect of jet quenching are presented; the first is the modification of spectrum of jet \pT in each IC class. The second is the  average distribution of the jet constituent (the particles clustered into the jet) $z$ values, where $z$ is the ratio of the constituent \pT to the total jet \pT: $z_\mathrm{T}^\mathrm{cons.}\equiv{}p_\mathrm{T}^\mathrm{cons.}/p_\mathrm{T}^\mathrm{jet}$. Both jet \pT and $z$ spectra resulting from using the FNOs are compared to results using \MUSIC.

\subsection{FNO Performance for Flow Measurements}

In the MC, the $x$-axis is always aligned along the impact parameter, so that
$\phi$'s origin is always in the MC's reaction plane. Therefore flow is
directly visible in the hadronic $\phi$ distribution. As stated, the hadrons used
are those at mid-rapidity ($|\eta|<1$) without a \pT cut.

Azimuthal distributions of hadrons in peripheral events are shown in
Fig.~\ref{fig:bulk_phi}~(a) for the following sets of events. The sets are
characterized by the IC generated by the MC, and how the QGP flow is evolved,
using either MUSIC or FNO. The distributions using FNOs are directly compared in
ratio to the truth, \MUSIC, values in Fig.~\ref{fig:bulk_phi}~(b).

\begin{itemize}
    \item $\FNOper(\ICper)$; \ie peripherally trained FNO on ICs from peripheral events
    \item $\FNOcen(\ICper)$. This tests the sensitivity of the flow measurements
    to FNOs trained using central events but used on ICs from peripheral events.
    \item $\FNOcenper(\ICper)$. This tests the performance of a FNO trained on
    both central and peripheral events on ICs from peripheral events.
    \item $\FNOper(\ICperspikey)$. This tests the sensitivity of FNO calculations to ICs from events with nucleon widths outside the range used to train the FNO.
    \item $\MUSIC_{40}^{60}$: the truth level data for peripheral events.
    \item ${}^{0.8\mathrm{nw}}\MUSIC_{40}^{60}$: the truth level data for peripheral events using the smaller nucleon width.
\end{itemize}

\doublefigure
{\input{bulk_phi.tex}}{\LabPhi}
{Spectra of hadron azimuthal distributions from MC simulation in which the QGP flow is simulated by FNOs and calculated by the PDE solver \MUSIC.}
{\input{bulk_phi_ratio.tex}}{\LabPhiRat}
{Ratios of azimuthal distributions from MC simulations in which the QGP flow is simulated by FNOs to those calculated using the PDE solver \MUSIC.}
{Hadron azimuthal distributions in peripheral MC events using FNOs and \MUSIC, indicating flow in the QGP. Centralities using in the MC events, and those to train the FNOs, are given in right side super- and subscripts. If other than \SI{1.12}{fm}, the nucleon widths are give in in left-side superscripts.}
{fig:bulk_phi}

\doublefigure
{\input{bulk_pt.tex}}{\LabBulkPt}
{Spectra of hadron \pT from MC simulations in which the QGP flow is simulated by FNOs and calculated by the PDF solver \MUSIC.}
{\input{bulk_pt_ratio.tex}}{\LabBulkPtRat}
{Ratios of hadron \pT spectra from MC simulations with QGP flow simulated by FNOs to those calculated the flow with the PDE solver \MUSIC}
{Measurements of bulk hadron \pT spectra in MC simulations. 
Centralities used for ICs and to train the FNOs are indicated by right side super-(sub-)scripts. The nucleon widths, if other than \SI{1.12}{fm}, are given the left-side superscripts.}
{fig:bulk_pt}

\doublefigure
{\input{fnoall_phi.tex}}{\LabPhi}
{ Azimuthal distributions of hadrons from peripheral MC simulations using three different nucleon widths, with QGP flow either predicted by an FNO (trained using four kinds of MC events) or calculated by \MUSIC.
}
{\input{fnoall_phi_ratio.tex}}{\LabPhiRat}
{ Ratios of azimuthal distributions of hadrons in MC simulations using FNOs to predict QGP flow to MC simulations calculating QGP flow using \MUSIC.} 
{Hadron azimuthal distributions in MC simulations of peripheral collisions (using three different nucleon widths) in which the QGP evolution is either predicted by an FNO trained on four different kinds of events, or calculated by \MUSIC.  Centralities used for ICs and to train the FNOs are indicated by right side super- and subscripts. The nucleon widths are given the left-side superscripts.}
{fig:FNOallspikes}

The results for $\FNOper(\ICper)$ and $\FNOcenper(\ICper)$ demonstrate that the FNOs trained in whole or part using IC from peripheral events are consistent with the truth values. The smaller nucleon widths in the MC result in stronger (elliptic) flow, as shown in the $dN/d\phi$ distribution of \MUSICperspikey, which is failed to be captured by the $\FNOper(\ICperspikey)$ prediction. It is also apparent from $\FNOcen(\ICper)$ that training only on central events yields a FNO that biases flow measurements in peripheral events.

Figure~\ref{fig:bulk_pt} presents the \pT distributions and ratios for the same data sets. This confirms the consistency of acceptability for \FNOper and \FNOcenper for evolving events with \ICper. Fig.~\ref{fig:bulk_pt}~(b) also show biases monotonically increasing in \pT, both for applying \FNOcen and \FNOper to \ICperspikey. In both cases, the FNOs under predict the production of higher \pT hadrons.

\subsection{Flow Measurement Performance for Events with Novel Nucleon Widths}

Based on the success of the $\FNOcenper(\ICper)$ model, we wanted to explore further if the FNO can interpolate properly to events with intermediate initial state nucleon widths not represented in the training events. Consequently, one additional FNO is trained, this time on only 2,000 events total; 500 events from each of \ICcen, \ICper, \ICcenspikey, and \ICperspikey. This operator (labeled \FNOall) is then tested on peripheral events using three different nucleon widths. The results of \FNOall on these ICs, two using the same nucleon widths, \SI{0.8}{fm} and \SI{1.12}{fm}, used to train the FNO, and one halfway between at \SI{0.96}{fm}, on azimuthal hadron distributions are shown in Fig.~\ref{fig:FNOallspikes}.

As discussed above, there is an event selection bias for the different nucleon widths introduced by discarding events which do not reach freeze out by \fmc{6.4}. Events with smaller nucleon widths tend to have regions of higher \erho values -- essentially hot spots -- that take longer to cool to the point of freeze out. This bias is apparent in Fig.~\ref{fig:FNOallspikes}. As seen in Fig.~\ref{fig:FNOallspikes}~(a), \FNOall captures this event selection bias. Additionally, as shown in Fig.~\ref{fig:FNOallspikes}~(b), the flow measurement performance for the FNO for ICs resulting from each nucleon width is in agreement with the truth values calculated using \MUSIC. This demonstrates an ability to extrapolate the applicability of FNOs to ICs trained in events using intermediary nucleon widths relative to those used in the training.



\singlefigure{\input{fnoall_bulk_pt_ratio.tex}}
{\LabBulkPtRat}
{Ratios of hadron \pT distributions resulting from QGP flow at time of freeze out for three sets of events which are identical except for nucleon width. The results use \FNOall, trained on 500 events each of the largest and smallest nucleon widths with peripheral events (for 1,000 ICs total), and then again for central events (for another 1,000 ICs). The ratios are the ratio of the FNO predicted values to the truth values calculated by \MUSIC.}
{fig:fnoall_pt_ratio}

The performance figures for \FNOall on flow have several interesting and important consequences for potential applications in large scale Bayesian analysis. First, they indicate that a relatively small number of events may be sufficient to train a model (thousands of various types rather than tens of thousands). Furthermore, they imply that mixing events with different conditions to train an FNO does not degrade the trained FNO's performance on events matching any of the individual training conditions. Perhaps the most consequential result is that, as stated above, an FNO may interpolate accurately to events with nucleon widths bounded by the ICs used in its training, but with which it has not been trained at all.

\begin{figure*}[htbp!]
    \begin{minipage}{1.00\linewidth}
        \begin{minipage}{0.48\linewidth}
            \resizebox{1.00\linewidth}{!}{\clipbox{0.3cm 0cm 1.7cm 0.0cm}{\input{jet_pt.tex}}}
            \begin{tikzpicture}[overlay, remember picture]
                \LabJetPt
            \end{tikzpicture}
        \end{minipage}
        \begin{minipage}{0.48\linewidth}
            \resizebox{1.0\linewidth}{!}{\clipbox{0.3cm 0cm 1.7cm 0cm}{\input{jet_pt_rat.tex}}}
            \begin{tikzpicture}[overlay, remember picture]
                \LabJetPtRat
            \end{tikzpicture}
        \end{minipage}\\
        \raggedright
        \hspace{1.1cm}
        \begin{minipage}[t]{0.42\linewidth}
            \justifying
            \hspace{-3em}(a) Spectra of jet-\pT in MC simulations in which QGP evolution is either predicted by FNOs or calculated by \MUSIC. The $pp$ spectrum shows the values in which the MC has no quenching.
        \end{minipage}
        \hspace{1.0cm}
        \begin{minipage}[t]{0.42\linewidth}
            \justifying
            \hspace{-3.3em} (b) Ratios of jet-\pT spectra in MC simulations predicting QGP evolution using FNOs to those in which the QGP is calculated using \MUSIC.
        \end{minipage}
    \end{minipage}
    \caption{\justifying Jet quenching measured in MC simulations through jet-\pT distributions. Centralities used for ICs and to train the FNOs are indicated by right side super-and subscripts. The nucleon widths are given the left-side superscripts.}\label{fig:jet_pt}
    \begin{minipage}{1.00\linewidth}
        \begin{minipage}{0.48\linewidth}
            \resizebox{1.00\linewidth}{!}{\clipbox{0.3cm 0cm 1.7cm 0.0cm}{\input{jet_z.tex}}}
            \begin{tikzpicture}[overlay, remember picture]
                \LabJetZ
            \end{tikzpicture}
        \end{minipage}
        \begin{minipage}{0.48\linewidth}
            \resizebox{1.0\linewidth}{!}{\clipbox{0.3cm 0cm 1.7cm 0cm}{\input{jet_zrat.tex}}}
            \begin{tikzpicture}[overlay, remember picture]
                \LabJetZrat
            \end{tikzpicture}
        \end{minipage}\\[-1em]
        \raggedright
        \hspace{1.1cm}
        \begin{minipage}[t]{0.42\linewidth}
            \justifying
            \hspace{-3em}(a) Fragmentation function distributions, $z$, for jets quenching in QGP in MC simulations where QGP evolution is either predicted by FNOs or calculated by \MUSIC.
        \end{minipage}
        \hspace{1.0cm}
        \begin{minipage}[t]{0.42\linewidth}
            \justifying
            \hspace{-3.3em} (b) Ratio of $z$ distributions from MCs using FNO predictions for QGP flow to those using calculations by \MUSIC.
        \end{minipage}
    \end{minipage}
    \caption{\justifying Jet quenching measured in MC simulations through $z$, the fractional \pT of jet constituents relative to jet \pT events. Centralities used for ICs and to train the FNOs are indicated by right side super- and subscripts. The nucleon widths, if other than \SI{1.12}{fm}, are given the left-side superscripts. For a second presentation of panel~(b), but without the \MUSIC/$pp$ ratio, see Fig.~\ref{fig:jet_zrat_zoomed} in the Appendix.}\label{fig:jet_z}
\end{figure*}


\subsection{FNO Performance for Jet Quenching}

The results of jet quenching in central events for the first \fmc{6.5}, using \SI{20}{GeV/\mathit{c}} gluons, are presented in Fig.s~\ref{fig:jet_pt}-\ref{fig:jet_z}, using central events with both the nominal and smaller nucleon width in the MC. The jet quenching observables are reported through modification to the jet \pT spectra and the jet fragmentation function as measured through the $z$ distributions of the constituent particles, where $z$ is defined as $z_\mathrm{const.}\equiv p_\mathrm{T,const.}/p_\mathrm{T,jet}$. An additional set of jets have been produced using identical MC settings but setting the QGP interaction length of QGP to $0$. This turns off jet quenching in the MC, and allows a direct comparison for the magnitude of quenching effects shown in  Fig.s~\ref{fig:jet_pt}-\ref{fig:jet_z}. As shown in a previous study, these jets are equivalent to jets generated in a MC using proton on proton collisions \cite{Stewart:2024mkx}, and as such are labeled $pp$ in the figures. For the jet quenching MC settings in \JETSCAPE used for this study, one can clearly see in the figures that with respect to the $pp$ baseline significant modification in the jet energy and in the fragmentation function are visible.

The values of \MUSICcen and \MUSICcenspikey in Fig.~\ref{fig:jet_z} and Fig.~\ref{fig:jet_pt} show that for the jets studied, jet quenching is not sensitive to the selection of nucleon width in the MC. This does not imply in general that jet quenching is insensitive to details in the initial conditions, rather that for our specific kinematic selection and observables we do not observe any difference. For convenience, the data in Fig.~\ref{fig:jet_z}~(b) is displayed again in the appendix in Fig.~\ref{fig:jet_zrat_zoomed}, in which the \MUSIC/pp ratio is removed and the range of the $y$-axis is decreased.

Fig.~\ref{fig:jet_z} show the $z$  distribution is strongly modified so that there are many more soft, low-$z$, and fewer high-$z$, constituents. The quenching is also reflected in the softer jet \pT distribution, relative to $pp$, shown in Fig.~\ref{fig:jet_pt}. The comparisons between the \MUSICcen, \FNOcen, and \FNOcenper, show that the FNOs performance for measuring jet quenching is consistent within statistical uncertainties to the \MUSIC truth and therefore, FNOs can be utilized as a fast and accurate alternative in jet quenching simulations.

\subsection{Benchmarking FNO Runtimes}

Some (rough) benchmarking results comparing the runtime of the FNO implementation in \JETSCAPE to the standard \MUSIC workflow are presented here. It is worth noting that there are mostly likely multiple optimizations possible beyond the simple implementations used in this study. For this comparison, FNO run times are measured using synthetic data; \ie of the proper format to test FNO inference times but not suitable for physics evaluations. The synthetic data is of the standard \JETSCAPE spatial resolution of $150\times150$ grid in \xy and a QGP evolution time of \fmc{12-15} (as opposed to the \fmc{6} trained in this study). Using the AMD system (see ~Sec.~\ref{app:hardware}) as well as an Apple M3 Max laptop, the FNO inference is 2-3 times faster than \MUSIC in central events when run using a single core on the CPU. \MUSIC's simulation time decreases with event size, while FNO inference times are constant regardless of event size. This negates FNO run time's advantage relative to \MUSIC in peripheral events. However, the real improvement lies in the ability to utilize GPU acceleration like CUDA or MPS on Apple silicon for FNO model inference. 

While it was not possible to train the FNO model for full central events on our GPU due to memory restrictions, using the PyTorch \texttt{torch::NoGradGuard} settings, suitable to test inference times only, we were able to test the FNO runtime performance with the above inputs on the NVIDIA GeForce RTX 3090. In this case, the FNO prediction of the full hydro evolution in central events took around 250 milliseconds, as compared to around 60 seconds for \MUSIC on a single CPU core (excluding any other overhead), resulting in an impressive reduction of computing time by a factor of more than 200. Utilizing Apple's MPS, we do not expect an improvement of the same factor, but conservatively, an order of magnitude can easily be achieved on Apple silicon computers\footnote{More detailed benchmarking results, including Apple silicon and other systems, will be provided and kept up-to-date on the GitHub repository \cite{githubarch}.}.


\section{Summary and Outlook}\label{sec:conclusions}

This study has presented the first study testing FNOs trained to model the QGP ultra-relativistic hydrodynamic flow in heavy ion MC events. The results show that the FNO performance is very good for collision events whose centrality and nucleon widths match those used to train the FNOs, but suffers when they do not. However, it is also shown that when an FNO is trained on multiple centralities and/or nucleon widths, then it appears to perform equally well for each class of events. Furthermore, FNOs trained on events with two nucleon widths also performs equally well for events using a nucleon width between the nucleon widths used for the training. This result is particularly interesting, as it implies that it may be possible to train a single FNO that will be applicable reproduce the hydrodynamic evolution of events with significant variation in multiple parameters characterizing the initial conditions, as has been successfully implemented for FNOs modeling classical hydrodynamics \cite{Li2021Fourier}. Such an FNO would prove particularly useful in very computational time consuming Bayesian analyses.

Due to the complexity of the concurrent components of heavy-ion MC, an FNO must be qualified for use with each specific physics observable.  This study has demonstrated that the FNO performance is very good in reproducing standard flow and jet quenching measurements. While the effect is not verified via a flow measurement, direct inspection of spatial super-resolution QGP evolution predictions, indicate the FNO performance at super-resolution is equal to its performance at resolutions matching the training resolutions. This feature may prove useful in future flow measurements.

These successes indicates that FNOs can be used to dramatically speed up MC simulations using QGP flow in our setup by more than a factor of 200 using GPU acceleration. The code used in this paper is archived at \cite{githubarch} in order to facilitate further studies required for more direct applications of NFOs in heavy-ion physics research. To that end, there are several promising topics of study to pursue.

With additional GPU memory capacity, via either multiple GPUs with more sophisticated training algorithms, or even simply a larger GPU, an FNO can be trained to a full time evolution for central collisions, instead of only the first \fmc{6}. Additionally, with more resources, an extension to train and use an FNO for 3+1D hydrodynamic evolution is anticipated to be straightforward. Additionally, with more memory, FNO modeling of viscous hydrodynamic flow, which increases the number of QGP parameters from three to nine, can be investigated.

Finally, it would be particularly useful if an FNO can be trained to propagate short distances in \ttau from arbitrary starting points, as has been successfully implemented for FNOs modeling classical hydrodynamics \cite{Li2021Fourier}. Such a FNO could be used to iteratively evolve flow to arbitrarily large \ttau values. It could also be used for MC models accounting for the concurrent interplay between jet energy deposition, jet wakes, and hydrodynamic QGP evolution. Such computationally demanding simulations would be timely as signals of jet wakes are currently a topic of particularly active interest.

\section{Acknowledgments}

This work was supported by the U.S. Department of Energy Office of Science, Office of Nuclear Physics under Award No. DE-FG02-92ER40713, and the National Science Foundation (NSF) within the framework of the JETSCAPE collaboration, under grant number OAC-2004571(CSSI:X-SCAPE). We are very grateful to Dr. Chun Shen of Wayne State University, and a primary author of \JETSCAPE's \MUSIC module, for many conversations and emails which aided in this study. His generous time and support has been instrumental to our endeavors.

\balance
\bibliographystyle{apsrev4-2}
\bibliography{main}

\begin{thebibliography}{42}%
\makeatletter
\providecommand \@ifxundefined [1]{%
 \@ifx{#1\undefined}
}%
\providecommand \@ifnum [1]{%
 \ifnum #1\expandafter \@firstoftwo
 \else \expandafter \@secondoftwo
 \fi
}%
\providecommand \@ifx [1]{%
 \ifx #1\expandafter \@firstoftwo
 \else \expandafter \@secondoftwo
 \fi
}%
\providecommand \natexlab [1]{#1}%
\providecommand \enquote  [1]{``#1''}%
\providecommand \bibnamefont  [1]{#1}%
\providecommand \bibfnamefont [1]{#1}%
\providecommand \citenamefont [1]{#1}%
\providecommand \href@noop [0]{\@secondoftwo}%
\providecommand \href [0]{\begingroup \@sanitize@url \@href}%
\providecommand \@href[1]{\@@startlink{#1}\@@href}%
\providecommand \@@href[1]{\endgroup#1\@@endlink}%
\providecommand \@sanitize@url [0]{\catcode `\\12\catcode `\$12\catcode `\&12\catcode `\#12\catcode `\^12\catcode `\_12\catcode `\%12\relax}%
\providecommand \@@startlink[1]{}%
\providecommand \@@endlink[0]{}%
\providecommand \url  [0]{\begingroup\@sanitize@url \@url }%
\providecommand \@url [1]{\endgroup\@href {#1}{\urlprefix }}%
\providecommand \urlprefix  [0]{URL }%
\providecommand \Eprint [0]{\href }%
\providecommand \doibase [0]{https://doi.org/}%
\providecommand \selectlanguage [0]{\@gobble}%
\providecommand \bibinfo  [0]{\@secondoftwo}%
\providecommand \bibfield  [0]{\@secondoftwo}%
\providecommand \translation [1]{[#1]}%
\providecommand \BibitemOpen [0]{}%
\providecommand \bibitemStop [0]{}%
\providecommand \bibitemNoStop [0]{.\EOS\space}%
\providecommand \EOS [0]{\spacefactor3000\relax}%
\providecommand \BibitemShut  [1]{\csname bibitem#1\endcsname}%
\let\auto@bib@innerbib\@empty
\bibitem [{\citenamefont {Arsene}\ \emph {et~al.}(2005)\citenamefont {Arsene} \emph {et~al.}}]{BRAHMS:2004adc}%
  \BibitemOpen
  \bibfield  {author} {\bibinfo {author} {\bibfnamefont {I.}~\bibnamefont {Arsene}} \emph {et~al.} (\bibinfo {collaboration} {BRAHMS}),\ }\href {https://doi.org/10.1016/j.nuclphysa.2005.02.130} {\bibfield  {journal} {\bibinfo  {journal} {Nucl. Phys. A}\ }\textbf {\bibinfo {volume} {757}},\ \bibinfo {pages} {1} (\bibinfo {year} {2005})},\ \Eprint {https://arxiv.org/abs/nucl-ex/0410020} {arXiv:nucl-ex/0410020} \BibitemShut {NoStop}%
\bibitem [{\citenamefont {Back}\ \emph {et~al.}(2005)\citenamefont {Back} \emph {et~al.}}]{PHOBOS:2004zne}%
  \BibitemOpen
  \bibfield  {author} {\bibinfo {author} {\bibfnamefont {B.~B.}\ \bibnamefont {Back}} \emph {et~al.} (\bibinfo {collaboration} {PHOBOS}),\ }\href {https://doi.org/10.1016/j.nuclphysa.2005.03.084} {\bibfield  {journal} {\bibinfo  {journal} {Nucl. Phys. A}\ }\textbf {\bibinfo {volume} {757}},\ \bibinfo {pages} {28} (\bibinfo {year} {2005})},\ \Eprint {https://arxiv.org/abs/nucl-ex/0410022} {arXiv:nucl-ex/0410022} \BibitemShut {NoStop}%
\bibitem [{\citenamefont {Adams}\ \emph {et~al.}(2005)\citenamefont {Adams} \emph {et~al.}}]{STAR:2005gfr}%
  \BibitemOpen
  \bibfield  {author} {\bibinfo {author} {\bibfnamefont {J.}~\bibnamefont {Adams}} \emph {et~al.} (\bibinfo {collaboration} {STAR}),\ }\href {https://doi.org/10.1016/j.nuclphysa.2005.03.085} {\bibfield  {journal} {\bibinfo  {journal} {Nucl. Phys. A}\ }\textbf {\bibinfo {volume} {757}},\ \bibinfo {pages} {102} (\bibinfo {year} {2005})},\ \Eprint {https://arxiv.org/abs/nucl-ex/0501009} {arXiv:nucl-ex/0501009} \BibitemShut {NoStop}%
\bibitem [{\citenamefont {Adcox}\ \emph {et~al.}(2005)\citenamefont {Adcox} \emph {et~al.}}]{PHENIX:2004vcz}%
  \BibitemOpen
  \bibfield  {author} {\bibinfo {author} {\bibfnamefont {K.}~\bibnamefont {Adcox}} \emph {et~al.} (\bibinfo {collaboration} {PHENIX}),\ }\href {https://doi.org/10.1016/j.nuclphysa.2005.03.086} {\bibfield  {journal} {\bibinfo  {journal} {Nucl. Phys. A}\ }\textbf {\bibinfo {volume} {757}},\ \bibinfo {pages} {184} (\bibinfo {year} {2005})},\ \Eprint {https://arxiv.org/abs/nucl-ex/0410003} {arXiv:nucl-ex/0410003} \BibitemShut {NoStop}%
\bibitem [{\citenamefont {Aad}\ \emph {et~al.}(2010)\citenamefont {Aad} \emph {et~al.}}]{ATLAS:2010isq}%
  \BibitemOpen
  \bibfield  {author} {\bibinfo {author} {\bibfnamefont {G.}~\bibnamefont {Aad}} \emph {et~al.} (\bibinfo {collaboration} {ATLAS}),\ }\href {https://doi.org/10.1103/PhysRevLett.105.252303} {\bibfield  {journal} {\bibinfo  {journal} {Phys. Rev. Lett.}\ }\textbf {\bibinfo {volume} {105}},\ \bibinfo {pages} {252303} (\bibinfo {year} {2010})},\ \Eprint {https://arxiv.org/abs/1011.6182} {arXiv:1011.6182 [hep-ex]} \BibitemShut {NoStop}%
\bibitem [{\citenamefont {Chatrchyan}\ \emph {et~al.}(2011)\citenamefont {Chatrchyan} \emph {et~al.}}]{CMS:2011iwn}%
  \BibitemOpen
  \bibfield  {author} {\bibinfo {author} {\bibfnamefont {S.}~\bibnamefont {Chatrchyan}} \emph {et~al.} (\bibinfo {collaboration} {CMS}),\ }\href {https://doi.org/10.1103/PhysRevC.84.024906} {\bibfield  {journal} {\bibinfo  {journal} {Phys. Rev. C}\ }\textbf {\bibinfo {volume} {84}},\ \bibinfo {pages} {024906} (\bibinfo {year} {2011})},\ \Eprint {https://arxiv.org/abs/1102.1957} {arXiv:1102.1957 [nucl-ex]} \BibitemShut {NoStop}%
\bibitem [{\citenamefont {Aamodt}\ \emph {et~al.}(2010)\citenamefont {Aamodt} \emph {et~al.}}]{ALICE:2010suc}%
  \BibitemOpen
  \bibfield  {author} {\bibinfo {author} {\bibfnamefont {K.}~\bibnamefont {Aamodt}} \emph {et~al.} (\bibinfo {collaboration} {ALICE}),\ }\href {https://doi.org/10.1103/PhysRevLett.105.252302} {\bibfield  {journal} {\bibinfo  {journal} {Phys. Rev. Lett.}\ }\textbf {\bibinfo {volume} {105}},\ \bibinfo {pages} {252302} (\bibinfo {year} {2010})},\ \Eprint {https://arxiv.org/abs/1011.3914} {arXiv:1011.3914 [nucl-ex]} \BibitemShut {NoStop}%
\bibitem [{\citenamefont {Busza}\ \emph {et~al.}(2018)\citenamefont {Busza}, \citenamefont {Rajagopal},\ and\ \citenamefont {van~der Schee}}]{Busza:2018rrf}%
  \BibitemOpen
  \bibfield  {author} {\bibinfo {author} {\bibfnamefont {W.}~\bibnamefont {Busza}}, \bibinfo {author} {\bibfnamefont {K.}~\bibnamefont {Rajagopal}},\ and\ \bibinfo {author} {\bibfnamefont {W.}~\bibnamefont {van~der Schee}},\ }\href {https://doi.org/10.1146/annurev-nucl-101917-020852} {\bibfield  {journal} {\bibinfo  {journal} {Ann. Rev. Nucl. Part. Sci.}\ }\textbf {\bibinfo {volume} {68}},\ \bibinfo {pages} {339} (\bibinfo {year} {2018})},\ \Eprint {https://arxiv.org/abs/1802.04801} {arXiv:1802.04801 [hep-ph]} \BibitemShut {NoStop}%
\bibitem [{\citenamefont {Niida}\ and\ \citenamefont {Miake}(2021)}]{Niida:2021wut}%
  \BibitemOpen
  \bibfield  {author} {\bibinfo {author} {\bibfnamefont {T.}~\bibnamefont {Niida}}\ and\ \bibinfo {author} {\bibfnamefont {Y.}~\bibnamefont {Miake}},\ }\href {https://doi.org/10.1007/s43673-021-00014-3} {\bibfield  {journal} {\bibinfo  {journal} {AAPPS Bull.}\ }\textbf {\bibinfo {volume} {31}},\ \bibinfo {pages} {12} (\bibinfo {year} {2021})},\ \Eprint {https://arxiv.org/abs/2104.11406} {arXiv:2104.11406 [nucl-ex]} \BibitemShut {NoStop}%
\bibitem [{\citenamefont {K\v{r}\'\i{}\v{z}kov\'a~Gajdo\v{s}ov\'a}(2021)}]{KrizkovaGajdosova:2020pxc}%
  \BibitemOpen
  \bibfield  {author} {\bibinfo {author} {\bibfnamefont {K.}~\bibnamefont {K\v{r}\'\i{}\v{z}kov\'a~Gajdo\v{s}ov\'a}},\ }\href {https://doi.org/10.1016/j.nuclphysa.2020.121802} {\bibfield  {journal} {\bibinfo  {journal} {Nucl. Phys. A}\ }\textbf {\bibinfo {volume} {1005}},\ \bibinfo {pages} {121802} (\bibinfo {year} {2021})},\ \Eprint {https://arxiv.org/abs/2007.12529} {arXiv:2007.12529 [nucl-ex]} \BibitemShut {NoStop}%
\bibitem [{\citenamefont {Cunqueiro}\ and\ \citenamefont {Sickles}(2022)}]{Cunqueiro:2021wls}%
  \BibitemOpen
  \bibfield  {author} {\bibinfo {author} {\bibfnamefont {L.}~\bibnamefont {Cunqueiro}}\ and\ \bibinfo {author} {\bibfnamefont {A.~M.}\ \bibnamefont {Sickles}},\ }\href {https://doi.org/10.1016/j.ppnp.2022.103940} {\bibfield  {journal} {\bibinfo  {journal} {Prog. Part. Nucl. Phys.}\ }\textbf {\bibinfo {volume} {124}},\ \bibinfo {pages} {103940} (\bibinfo {year} {2022})},\ \Eprint {https://arxiv.org/abs/2110.14490} {arXiv:2110.14490 [nucl-ex]} \BibitemShut {NoStop}%
\bibitem [{\citenamefont {Eskola}\ \emph {et~al.}(2017)\citenamefont {Eskola}, \citenamefont {Paakkinen}, \citenamefont {Paukkunen},\ and\ \citenamefont {Salgado}}]{Eskola:2016oht}%
  \BibitemOpen
  \bibfield  {author} {\bibinfo {author} {\bibfnamefont {K.~J.}\ \bibnamefont {Eskola}}, \bibinfo {author} {\bibfnamefont {P.}~\bibnamefont {Paakkinen}}, \bibinfo {author} {\bibfnamefont {H.}~\bibnamefont {Paukkunen}},\ and\ \bibinfo {author} {\bibfnamefont {C.~A.}\ \bibnamefont {Salgado}},\ }\href {https://doi.org/10.1140/epjc/s10052-017-4725-9} {\bibfield  {journal} {\bibinfo  {journal} {Eur. Phys. J. C}\ }\textbf {\bibinfo {volume} {77}},\ \bibinfo {pages} {163} (\bibinfo {year} {2017})},\ \Eprint {https://arxiv.org/abs/1612.05741} {arXiv:1612.05741 [hep-ph]} \BibitemShut {NoStop}%
\bibitem [{\citenamefont {Miller}\ \emph {et~al.}(2007)\citenamefont {Miller}, \citenamefont {Reygers}, \citenamefont {Sanders},\ and\ \citenamefont {Steinberg}}]{Miller:2007ri}%
  \BibitemOpen
  \bibfield  {author} {\bibinfo {author} {\bibfnamefont {M.~L.}\ \bibnamefont {Miller}}, \bibinfo {author} {\bibfnamefont {K.}~\bibnamefont {Reygers}}, \bibinfo {author} {\bibfnamefont {S.~J.}\ \bibnamefont {Sanders}},\ and\ \bibinfo {author} {\bibfnamefont {P.}~\bibnamefont {Steinberg}},\ }\href {https://doi.org/10.1146/annurev.nucl.57.090506.123020} {\bibfield  {journal} {\bibinfo  {journal} {Ann. Rev. Nucl. Part. Sci.}\ }\textbf {\bibinfo {volume} {57}},\ \bibinfo {pages} {205} (\bibinfo {year} {2007})},\ \Eprint {https://arxiv.org/abs/nucl-ex/0701025} {arXiv:nucl-ex/0701025} \BibitemShut {NoStop}%
\bibitem [{\citenamefont {Krasnitz}\ and\ \citenamefont {Venugopalan}(2000)}]{Krasnitz:1999wc}%
  \BibitemOpen
  \bibfield  {author} {\bibinfo {author} {\bibfnamefont {A.}~\bibnamefont {Krasnitz}}\ and\ \bibinfo {author} {\bibfnamefont {R.}~\bibnamefont {Venugopalan}},\ }\href {https://doi.org/10.1103/PhysRevLett.84.4309} {\bibfield  {journal} {\bibinfo  {journal} {Phys. Rev. Lett.}\ }\textbf {\bibinfo {volume} {84}},\ \bibinfo {pages} {4309} (\bibinfo {year} {2000})},\ \Eprint {https://arxiv.org/abs/hep-ph/9909203} {arXiv:hep-ph/9909203} \BibitemShut {NoStop}%
\bibitem [{\citenamefont {Shen}\ and\ \citenamefont {Schenke}(2018)}]{Shen:2017bsr}%
  \BibitemOpen
  \bibfield  {author} {\bibinfo {author} {\bibfnamefont {C.}~\bibnamefont {Shen}}\ and\ \bibinfo {author} {\bibfnamefont {B.}~\bibnamefont {Schenke}},\ }\href {https://doi.org/10.1103/PhysRevC.97.024907} {\bibfield  {journal} {\bibinfo  {journal} {Phys. Rev. C}\ }\textbf {\bibinfo {volume} {97}},\ \bibinfo {pages} {024907} (\bibinfo {year} {2018})},\ \Eprint {https://arxiv.org/abs/1710.00881} {arXiv:1710.00881 [nucl-th]} \BibitemShut {NoStop}%
\bibitem [{\citenamefont {Fries}\ \emph {et~al.}(2025)\citenamefont {Fries}, \citenamefont {Greco},\ and\ \citenamefont {Rapp}}]{Fries:2025jfi}%
  \BibitemOpen
  \bibfield  {author} {\bibinfo {author} {\bibfnamefont {R.~J.}\ \bibnamefont {Fries}}, \bibinfo {author} {\bibfnamefont {V.}~\bibnamefont {Greco}},\ and\ \bibinfo {author} {\bibfnamefont {R.}~\bibnamefont {Rapp}},\ }\href@noop {} {\  (\bibinfo {year} {2025})},\ \Eprint {https://arxiv.org/abs/2506.24023} {arXiv:2506.24023 [hep-ph]} \BibitemShut {NoStop}%
\bibitem [{\citenamefont {Altmann}\ \emph {et~al.}(2025)\citenamefont {Altmann}, \citenamefont {Dubla}, \citenamefont {Greco}, \citenamefont {Rossi},\ and\ \citenamefont {Skands}}]{Altmann:2024kwx}%
  \BibitemOpen
  \bibfield  {author} {\bibinfo {author} {\bibfnamefont {J.}~\bibnamefont {Altmann}}, \bibinfo {author} {\bibfnamefont {A.}~\bibnamefont {Dubla}}, \bibinfo {author} {\bibfnamefont {V.}~\bibnamefont {Greco}}, \bibinfo {author} {\bibfnamefont {A.}~\bibnamefont {Rossi}},\ and\ \bibinfo {author} {\bibfnamefont {P.}~\bibnamefont {Skands}},\ }\href {https://doi.org/10.1140/epjc/s10052-024-13641-5} {\bibfield  {journal} {\bibinfo  {journal} {Eur. Phys. J. C}\ }\textbf {\bibinfo {volume} {85}},\ \bibinfo {pages} {16} (\bibinfo {year} {2025})},\ \Eprint {https://arxiv.org/abs/2405.19137} {arXiv:2405.19137 [hep-ph]} \BibitemShut {NoStop}%
\bibitem [{\citenamefont {Putschke}\ \emph {et~al.}(2019)\citenamefont {Putschke} \emph {et~al.}}]{Putschke:2019yrg}%
  \BibitemOpen
  \bibfield  {author} {\bibinfo {author} {\bibfnamefont {J.~H.}\ \bibnamefont {Putschke}} \emph {et~al.},\ }\href@noop {} {\  (\bibinfo {year} {2019})},\ \Eprint {https://arxiv.org/abs/1903.07706} {arXiv:1903.07706 [nucl-th]} \BibitemShut {NoStop}%
\bibitem [{\citenamefont {Everett}\ \emph {et~al.}(2021)\citenamefont {Everett} \emph {et~al.}}]{JETSCAPE:2020mzn}%
  \BibitemOpen
  \bibfield  {author} {\bibinfo {author} {\bibfnamefont {D.}~\bibnamefont {Everett}} \emph {et~al.} (\bibinfo {collaboration} {JETSCAPE}),\ }\href {https://doi.org/10.1103/PhysRevC.103.054904} {\bibfield  {journal} {\bibinfo  {journal} {Phys. Rev. C}\ }\textbf {\bibinfo {volume} {103}},\ \bibinfo {pages} {054904} (\bibinfo {year} {2021})},\ \Eprint {https://arxiv.org/abs/2011.01430} {arXiv:2011.01430 [hep-ph]} \BibitemShut {NoStop}%
\bibitem [{\citenamefont {Shen}\ and\ \citenamefont {Yan}(2020)}]{Shen:2020mgh}%
  \BibitemOpen
  \bibfield  {author} {\bibinfo {author} {\bibfnamefont {C.}~\bibnamefont {Shen}}\ and\ \bibinfo {author} {\bibfnamefont {L.}~\bibnamefont {Yan}},\ }\href {https://doi.org/10.1007/s41365-020-00829-z} {\bibfield  {journal} {\bibinfo  {journal} {Nucl. Sci. Tech.}\ }\textbf {\bibinfo {volume} {31}},\ \bibinfo {pages} {122} (\bibinfo {year} {2020})},\ \Eprint {https://arxiv.org/abs/2010.12377} {arXiv:2010.12377 [nucl-th]} \BibitemShut {NoStop}%
\bibitem [{\citenamefont {Heinz}(2005)}]{Heinz:2005zg}%
  \BibitemOpen
  \bibfield  {author} {\bibinfo {author} {\bibfnamefont {U.~W.}\ \bibnamefont {Heinz}},\ }in\ \href@noop {} {\emph {\bibinfo {booktitle} {{Workshop on Extreme QCD}}}}\ (\bibinfo {year} {2005})\ pp.\ \bibinfo {pages} {3--12},\ \Eprint {https://arxiv.org/abs/nucl-th/0512051} {arXiv:nucl-th/0512051} \BibitemShut {NoStop}%
\bibitem [{\citenamefont {Schenke}\ \emph {et~al.}(2010)\citenamefont {Schenke}, \citenamefont {Jeon},\ and\ \citenamefont {Gale}}]{Schenke:2010nt}%
  \BibitemOpen
  \bibfield  {author} {\bibinfo {author} {\bibfnamefont {B.}~\bibnamefont {Schenke}}, \bibinfo {author} {\bibfnamefont {S.}~\bibnamefont {Jeon}},\ and\ \bibinfo {author} {\bibfnamefont {C.}~\bibnamefont {Gale}},\ }\href {https://doi.org/10.1103/PhysRevC.82.014903} {\bibfield  {journal} {\bibinfo  {journal} {Phys. Rev. C}\ }\textbf {\bibinfo {volume} {82}},\ \bibinfo {pages} {014903} (\bibinfo {year} {2010})},\ \Eprint {https://arxiv.org/abs/1004.1408} {arXiv:1004.1408 [hep-ph]} \BibitemShut {NoStop}%
\bibitem [{\citenamefont {Schenke}\ \emph {et~al.}(2011)\citenamefont {Schenke}, \citenamefont {Jeon},\ and\ \citenamefont {Gale}}]{Schenke:2010rr}%
  \BibitemOpen
  \bibfield  {author} {\bibinfo {author} {\bibfnamefont {B.}~\bibnamefont {Schenke}}, \bibinfo {author} {\bibfnamefont {S.}~\bibnamefont {Jeon}},\ and\ \bibinfo {author} {\bibfnamefont {C.}~\bibnamefont {Gale}},\ }\href {https://doi.org/10.1103/PhysRevLett.106.042301} {\bibfield  {journal} {\bibinfo  {journal} {Phys. Rev. Lett.}\ }\textbf {\bibinfo {volume} {106}},\ \bibinfo {pages} {042301} (\bibinfo {year} {2011})},\ \Eprint {https://arxiv.org/abs/1009.3244} {arXiv:1009.3244 [hep-ph]} \BibitemShut {NoStop}%
\bibitem [{\citenamefont {Paquet}\ \emph {et~al.}(2016)\citenamefont {Paquet}, \citenamefont {Shen}, \citenamefont {Denicol}, \citenamefont {Luzum}, \citenamefont {Schenke}, \citenamefont {Jeon},\ and\ \citenamefont {Gale}}]{Paquet:2015lta}%
  \BibitemOpen
  \bibfield  {author} {\bibinfo {author} {\bibfnamefont {J.-F.}\ \bibnamefont {Paquet}}, \bibinfo {author} {\bibfnamefont {C.}~\bibnamefont {Shen}}, \bibinfo {author} {\bibfnamefont {G.~S.}\ \bibnamefont {Denicol}}, \bibinfo {author} {\bibfnamefont {M.}~\bibnamefont {Luzum}}, \bibinfo {author} {\bibfnamefont {B.}~\bibnamefont {Schenke}}, \bibinfo {author} {\bibfnamefont {S.}~\bibnamefont {Jeon}},\ and\ \bibinfo {author} {\bibfnamefont {C.}~\bibnamefont {Gale}},\ }\href {https://doi.org/10.1103/PhysRevC.93.044906} {\bibfield  {journal} {\bibinfo  {journal} {Phys. Rev. C}\ }\textbf {\bibinfo {volume} {93}},\ \bibinfo {pages} {044906} (\bibinfo {year} {2016})},\ \Eprint {https://arxiv.org/abs/1509.06738} {arXiv:1509.06738 [hep-ph]} \BibitemShut {NoStop}%
\bibitem [{\citenamefont {Kovachki}\ \emph {et~al.}(2023)\citenamefont {Kovachki}, \citenamefont {Li}, \citenamefont {Liu}, \citenamefont {Azizzadenesheli}, \citenamefont {Bhattacharya}, \citenamefont {Stuart},\ and\ \citenamefont {Anandkumar}}]{JMLR:v24:21-1524}%
  \BibitemOpen
  \bibfield  {author} {\bibinfo {author} {\bibfnamefont {N.}~\bibnamefont {Kovachki}}, \bibinfo {author} {\bibfnamefont {Z.}~\bibnamefont {Li}}, \bibinfo {author} {\bibfnamefont {B.}~\bibnamefont {Liu}}, \bibinfo {author} {\bibfnamefont {K.}~\bibnamefont {Azizzadenesheli}}, \bibinfo {author} {\bibfnamefont {K.}~\bibnamefont {Bhattacharya}}, \bibinfo {author} {\bibfnamefont {A.}~\bibnamefont {Stuart}},\ and\ \bibinfo {author} {\bibfnamefont {A.}~\bibnamefont {Anandkumar}},\ }\href {http://jmlr.org/papers/v24/21-1524.html} {\bibfield  {journal} {\bibinfo  {journal} {Journal of Machine Learning Research}\ }\textbf {\bibinfo {volume} {24}},\ \bibinfo {pages} {1} (\bibinfo {year} {2023})}\BibitemShut {NoStop}%
\bibitem [{\citenamefont {Azizzadenesheli}\ \emph {et~al.}(2024)\citenamefont {Azizzadenesheli}, \citenamefont {Kovachki}, \citenamefont {Li}, \citenamefont {Liu-Schiaffini}, \citenamefont {Kossaifi},\ and\ \citenamefont {Anandkumar}}]{Azizzadenesheli2024}%
  \BibitemOpen
  \bibfield  {author} {\bibinfo {author} {\bibfnamefont {K.}~\bibnamefont {Azizzadenesheli}}, \bibinfo {author} {\bibfnamefont {N.}~\bibnamefont {Kovachki}}, \bibinfo {author} {\bibfnamefont {Z.}~\bibnamefont {Li}}, \bibinfo {author} {\bibfnamefont {M.}~\bibnamefont {Liu-Schiaffini}}, \bibinfo {author} {\bibfnamefont {J.}~\bibnamefont {Kossaifi}},\ and\ \bibinfo {author} {\bibfnamefont {A.}~\bibnamefont {Anandkumar}},\ }\href@noop {} {\bibfield  {journal} {\bibinfo  {journal} {Nature Reviews Physics}\ }\textbf {\bibinfo {volume} {6}},\ \bibinfo {pages} {123–135} (\bibinfo {year} {2024})}\BibitemShut {NoStop}%
\bibitem [{\citenamefont {Kossaifi}\ \emph {et~al.}(2025)\citenamefont {Kossaifi}, \citenamefont {Kovachki}, \citenamefont {Li}, \citenamefont {Pitt}, \citenamefont {Liu-Schiaffini}, \citenamefont {George}, \citenamefont {Bonev}, \citenamefont {Azizzadenesheli}, \citenamefont {Berner}, \citenamefont {Duruisseaux},\ and\ \citenamefont {Anandkumar}}]{kossaifi2025librarylearningneuraloperators}%
  \BibitemOpen
  \bibfield  {author} {\bibinfo {author} {\bibfnamefont {J.}~\bibnamefont {Kossaifi}}, \bibinfo {author} {\bibfnamefont {N.}~\bibnamefont {Kovachki}}, \bibinfo {author} {\bibfnamefont {Z.}~\bibnamefont {Li}}, \bibinfo {author} {\bibfnamefont {D.}~\bibnamefont {Pitt}}, \bibinfo {author} {\bibfnamefont {M.}~\bibnamefont {Liu-Schiaffini}}, \bibinfo {author} {\bibfnamefont {R.~J.}\ \bibnamefont {George}}, \bibinfo {author} {\bibfnamefont {B.}~\bibnamefont {Bonev}}, \bibinfo {author} {\bibfnamefont {K.}~\bibnamefont {Azizzadenesheli}}, \bibinfo {author} {\bibfnamefont {J.}~\bibnamefont {Berner}}, \bibinfo {author} {\bibfnamefont {V.}~\bibnamefont {Duruisseaux}},\ and\ \bibinfo {author} {\bibfnamefont {A.}~\bibnamefont {Anandkumar}},\ }\href {https://arxiv.org/abs/2412.10354} {\bibinfo {title} {A library for learning neural operators}} (\bibinfo {year} {2025}),\ \Eprint {https://arxiv.org/abs/2412.10354} {arXiv:2412.10354 [cs.LG]} \BibitemShut {NoStop}%
\bibitem [{\citenamefont {Li}\ \emph {et~al.}(2021{\natexlab{a}})\citenamefont {Li}, \citenamefont {Kovachki}, \citenamefont {Azizzadenesheli}, \citenamefont {Liu}, \citenamefont {Bhattacharya}, \citenamefont {Stuart},\ and\ \citenamefont {Anandkumar}}]{li2021fourierneuraloperatorparametric}%
  \BibitemOpen
  \bibfield  {author} {\bibinfo {author} {\bibfnamefont {Z.}~\bibnamefont {Li}}, \bibinfo {author} {\bibfnamefont {N.}~\bibnamefont {Kovachki}}, \bibinfo {author} {\bibfnamefont {K.}~\bibnamefont {Azizzadenesheli}}, \bibinfo {author} {\bibfnamefont {B.}~\bibnamefont {Liu}}, \bibinfo {author} {\bibfnamefont {K.}~\bibnamefont {Bhattacharya}}, \bibinfo {author} {\bibfnamefont {A.}~\bibnamefont {Stuart}},\ and\ \bibinfo {author} {\bibfnamefont {A.}~\bibnamefont {Anandkumar}},\ }\href {https://arxiv.org/abs/2010.08895} {\bibinfo {title} {Fourier neural operator for parametric partial differential equations}} (\bibinfo {year} {2021}{\natexlab{a}}),\ \Eprint {https://arxiv.org/abs/2010.08895} {arXiv:2010.08895 [cs.LG]} \BibitemShut {NoStop}%
\bibitem [{\citenamefont {Li}\ \emph {et~al.}(2021{\natexlab{b}})\citenamefont {Li}, \citenamefont {Kovachki}, \citenamefont {Azizzadenesheli}, \citenamefont {Liu}, \citenamefont {Bhattacharya}, \citenamefont {Stuart},\ and\ \citenamefont {Anandkumar}}]{Li2021Fourier}%
  \BibitemOpen
  \bibfield  {author} {\bibinfo {author} {\bibfnamefont {Z.}~\bibnamefont {Li}}, \bibinfo {author} {\bibfnamefont {N.}~\bibnamefont {Kovachki}}, \bibinfo {author} {\bibfnamefont {K.}~\bibnamefont {Azizzadenesheli}}, \bibinfo {author} {\bibfnamefont {B.}~\bibnamefont {Liu}}, \bibinfo {author} {\bibfnamefont {K.}~\bibnamefont {Bhattacharya}}, \bibinfo {author} {\bibfnamefont {A.}~\bibnamefont {Stuart}},\ and\ \bibinfo {author} {\bibfnamefont {A.}~\bibnamefont {Anandkumar}},\ }\href {https://arxiv.org/abs/2108.08481v6} {\bibfield  {journal} {\bibinfo  {journal} {arXiv preprint arXiv:2108.08481v6}\ } (\bibinfo {year} {2021}{\natexlab{b}})}\BibitemShut {NoStop}%
\bibitem [{\citenamefont {Pathak}\ \emph {et~al.}(2022)\citenamefont {Pathak}, \citenamefont {Subramanian}, \citenamefont {Harrington}, \citenamefont {Raja}, \citenamefont {Chattopadhyay}, \citenamefont {Mardani}, \citenamefont {Kurth}, \citenamefont {Hall}, \citenamefont {Li}, \citenamefont {Azizzadenesheli}, \citenamefont {Hassanzadeh}, \citenamefont {Kashinath},\ and\ \citenamefont {Anandkumar}}]{pathak2022fourcastnetglobaldatadrivenhighresolution}%
  \BibitemOpen
  \bibfield  {author} {\bibinfo {author} {\bibfnamefont {J.}~\bibnamefont {Pathak}}, \bibinfo {author} {\bibfnamefont {S.}~\bibnamefont {Subramanian}}, \bibinfo {author} {\bibfnamefont {P.}~\bibnamefont {Harrington}}, \bibinfo {author} {\bibfnamefont {S.}~\bibnamefont {Raja}}, \bibinfo {author} {\bibfnamefont {A.}~\bibnamefont {Chattopadhyay}}, \bibinfo {author} {\bibfnamefont {M.}~\bibnamefont {Mardani}}, \bibinfo {author} {\bibfnamefont {T.}~\bibnamefont {Kurth}}, \bibinfo {author} {\bibfnamefont {D.}~\bibnamefont {Hall}}, \bibinfo {author} {\bibfnamefont {Z.}~\bibnamefont {Li}}, \bibinfo {author} {\bibfnamefont {K.}~\bibnamefont {Azizzadenesheli}}, \bibinfo {author} {\bibfnamefont {P.}~\bibnamefont {Hassanzadeh}}, \bibinfo {author} {\bibfnamefont {K.}~\bibnamefont {Kashinath}},\ and\ \bibinfo {author} {\bibfnamefont {A.}~\bibnamefont {Anandkumar}},\ }\href {https://arxiv.org/abs/2202.11214} {\bibinfo {title} {Fourcastnet: A global data-driven high-resolution weather model using adaptive fourier neural
  operators}} (\bibinfo {year} {2022}),\ \Eprint {https://arxiv.org/abs/2202.11214} {arXiv:2202.11214 [physics.ao-ph]} \BibitemShut {NoStop}%
\bibitem [{\citenamefont {Akiba}\ \emph {et~al.}(2016)\citenamefont {Akiba} \emph {et~al.}}]{LHCForwardPhysicsWorkingGroup:2016ote}%
  \BibitemOpen
  \bibfield  {author} {\bibinfo {author} {\bibfnamefont {K.}~\bibnamefont {Akiba}} \emph {et~al.} (\bibinfo {collaboration} {LHC Forward Physics Working Group}),\ }\href {https://doi.org/10.1088/0954-3899/43/11/110201} {\bibfield  {journal} {\bibinfo  {journal} {J. Phys. G}\ }\textbf {\bibinfo {volume} {43}},\ \bibinfo {pages} {110201} (\bibinfo {year} {2016})},\ \Eprint {https://arxiv.org/abs/1611.05079} {arXiv:1611.05079 [hep-ph]} \BibitemShut {NoStop}%
\bibitem [{\citenamefont {Hentschinski}\ \emph {et~al.}(2023)\citenamefont {Hentschinski} \emph {et~al.}}]{Hentschinski:2022xnd}%
  \BibitemOpen
  \bibfield  {author} {\bibinfo {author} {\bibfnamefont {M.}~\bibnamefont {Hentschinski}} \emph {et~al.},\ }\href {https://doi.org/10.5506/APhysPolB.54.3-A2} {\bibfield  {journal} {\bibinfo  {journal} {Acta Phys. Polon. B}\ }\textbf {\bibinfo {volume} {54}},\ \bibinfo {pages} {3} (\bibinfo {year} {2023})},\ \Eprint {https://arxiv.org/abs/2203.08129} {arXiv:2203.08129 [hep-ph]} \BibitemShut {NoStop}%
\bibitem [{git()}]{githubarch}%
  \BibitemOpen
  \href@noop {} {}\bibinfo {note} {Unique code for this study are archived at \texttt{https://github.com/jhputschke/JETSCAPE-FNO}}\BibitemShut {NoStop}%
\bibitem [{\citenamefont {Shen}\ \emph {et~al.}(2016)\citenamefont {Shen}, \citenamefont {Qiu}, \citenamefont {Song}, \citenamefont {Bernhard}, \citenamefont {Bass},\ and\ \citenamefont {Heinz}}]{Shen:2014vra}%
  \BibitemOpen
  \bibfield  {author} {\bibinfo {author} {\bibfnamefont {C.}~\bibnamefont {Shen}}, \bibinfo {author} {\bibfnamefont {Z.}~\bibnamefont {Qiu}}, \bibinfo {author} {\bibfnamefont {H.}~\bibnamefont {Song}}, \bibinfo {author} {\bibfnamefont {J.}~\bibnamefont {Bernhard}}, \bibinfo {author} {\bibfnamefont {S.}~\bibnamefont {Bass}},\ and\ \bibinfo {author} {\bibfnamefont {U.}~\bibnamefont {Heinz}},\ }\href {https://doi.org/10.1016/j.cpc.2015.08.039} {\bibfield  {journal} {\bibinfo  {journal} {Comput. Phys. Commun.}\ }\textbf {\bibinfo {volume} {199}},\ \bibinfo {pages} {61} (\bibinfo {year} {2016})},\ \Eprint {https://arxiv.org/abs/1409.8164} {arXiv:1409.8164 [nucl-th]} \BibitemShut {NoStop}%
\bibitem [{\citenamefont {Denicol}\ \emph {et~al.}(2018)\citenamefont {Denicol}, \citenamefont {Gale}, \citenamefont {Jeon}, \citenamefont {Monnai}, \citenamefont {Schenke},\ and\ \citenamefont {Shen}}]{Denicol:2018wdp}%
  \BibitemOpen
  \bibfield  {author} {\bibinfo {author} {\bibfnamefont {G.~S.}\ \bibnamefont {Denicol}}, \bibinfo {author} {\bibfnamefont {C.}~\bibnamefont {Gale}}, \bibinfo {author} {\bibfnamefont {S.}~\bibnamefont {Jeon}}, \bibinfo {author} {\bibfnamefont {A.}~\bibnamefont {Monnai}}, \bibinfo {author} {\bibfnamefont {B.}~\bibnamefont {Schenke}},\ and\ \bibinfo {author} {\bibfnamefont {C.}~\bibnamefont {Shen}},\ }\href {https://doi.org/10.1103/PhysRevC.98.034916} {\bibfield  {journal} {\bibinfo  {journal} {Phys. Rev. C}\ }\textbf {\bibinfo {volume} {98}},\ \bibinfo {pages} {034916} (\bibinfo {year} {2018})},\ \Eprint {https://arxiv.org/abs/1804.10557} {arXiv:1804.10557 [nucl-th]} \BibitemShut {NoStop}%
\bibitem [{\citenamefont {Apolin\'ario}\ \emph {et~al.}(2022)\citenamefont {Apolin\'ario}, \citenamefont {Lee},\ and\ \citenamefont {Winn}}]{Apolinario:2022vzg}%
  \BibitemOpen
  \bibfield  {author} {\bibinfo {author} {\bibfnamefont {L.}~\bibnamefont {Apolin\'ario}}, \bibinfo {author} {\bibfnamefont {Y.-J.}\ \bibnamefont {Lee}},\ and\ \bibinfo {author} {\bibfnamefont {M.}~\bibnamefont {Winn}},\ }\href {https://doi.org/10.1016/j.ppnp.2022.103990} {\bibfield  {journal} {\bibinfo  {journal} {Prog. Part. Nucl. Phys.}\ }\textbf {\bibinfo {volume} {127}},\ \bibinfo {pages} {103990} (\bibinfo {year} {2022})},\ \Eprint {https://arxiv.org/abs/2203.16352} {arXiv:2203.16352 [hep-ph]} \BibitemShut {NoStop}%
\bibitem [{\citenamefont {Sun}\ \emph {et~al.}(2024)\citenamefont {Sun}, \citenamefont {Wang}, \citenamefont {Ko}, \citenamefont {Ma},\ and\ \citenamefont {Shen}}]{Sun:2022xjr}%
  \BibitemOpen
  \bibfield  {author} {\bibinfo {author} {\bibfnamefont {K.-J.}\ \bibnamefont {Sun}}, \bibinfo {author} {\bibfnamefont {R.}~\bibnamefont {Wang}}, \bibinfo {author} {\bibfnamefont {C.~M.}\ \bibnamefont {Ko}}, \bibinfo {author} {\bibfnamefont {Y.-G.}\ \bibnamefont {Ma}},\ and\ \bibinfo {author} {\bibfnamefont {C.}~\bibnamefont {Shen}},\ }\href {https://doi.org/10.1038/s41467-024-45474-x} {\bibfield  {journal} {\bibinfo  {journal} {Nature Commun.}\ }\textbf {\bibinfo {volume} {15}},\ \bibinfo {pages} {1074} (\bibinfo {year} {2024})},\ \Eprint {https://arxiv.org/abs/2207.12532} {arXiv:2207.12532 [nucl-th]} \BibitemShut {NoStop}%
\bibitem [{\citenamefont {Nijs}\ and\ \citenamefont {van~der Schee}(2022)}]{Nijs:2022rme}%
  \BibitemOpen
  \bibfield  {author} {\bibinfo {author} {\bibfnamefont {G.}~\bibnamefont {Nijs}}\ and\ \bibinfo {author} {\bibfnamefont {W.}~\bibnamefont {van~der Schee}},\ }\href {https://doi.org/10.1103/PhysRevLett.129.232301} {\bibfield  {journal} {\bibinfo  {journal} {Phys. Rev. Lett.}\ }\textbf {\bibinfo {volume} {129}},\ \bibinfo {pages} {232301} (\bibinfo {year} {2022})},\ \Eprint {https://arxiv.org/abs/2206.13522} {arXiv:2206.13522 [nucl-th]} \BibitemShut {NoStop}%
\bibitem [{\citenamefont {Loshchilov}\ and\ \citenamefont {Hutter}(2019)}]{loshchilov2019decoupledweightdecayregularization}%
  \BibitemOpen
  \bibfield  {author} {\bibinfo {author} {\bibfnamefont {I.}~\bibnamefont {Loshchilov}}\ and\ \bibinfo {author} {\bibfnamefont {F.}~\bibnamefont {Hutter}},\ }\href {https://arxiv.org/abs/1711.05101} {\bibinfo {title} {Decoupled weight decay regularization}} (\bibinfo {year} {2019}),\ \Eprint {https://arxiv.org/abs/1711.05101} {arXiv:1711.05101 [cs.LG]} \BibitemShut {NoStop}%
\bibitem [{\citenamefont {Cacciari}\ \emph {et~al.}(2010)\citenamefont {Cacciari}, \citenamefont {Salam},\ and\ \citenamefont {Soyez}}]{Cacciari:2008gp}%
  \BibitemOpen
  \bibfield  {author} {\bibinfo {author} {\bibfnamefont {M.}~\bibnamefont {Cacciari}}, \bibinfo {author} {\bibfnamefont {G.~P.}\ \bibnamefont {Salam}},\ and\ \bibinfo {author} {\bibfnamefont {G.}~\bibnamefont {Soyez}},\ }\href {https://doi.org/10.1088/1126-6708/2008/04/063} {\bibfield  {journal} {\bibinfo  {journal} {JHEP}\ }\textbf {\bibinfo {volume} {04}},\ \bibinfo {pages} {063}},\ \Eprint {https://arxiv.org/abs/0802.1189} {arXiv:0802.1189 [hep-ph]} \BibitemShut {NoStop}%
\bibitem [{\citenamefont {Stewart}\ and\ \citenamefont {Putschke}(2025)}]{Stewart:2024mkx}%
  \BibitemOpen
  \bibfield  {author} {\bibinfo {author} {\bibfnamefont {D.}~\bibnamefont {Stewart}}\ and\ \bibinfo {author} {\bibfnamefont {J.}~\bibnamefont {Putschke}},\ }\href {https://doi.org/10.1103/PhysRevC.111.054902} {\bibfield  {journal} {\bibinfo  {journal} {Phys. Rev. C}\ }\textbf {\bibinfo {volume} {111}},\ \bibinfo {pages} {054902} (\bibinfo {year} {2025})},\ \Eprint {https://arxiv.org/abs/2412.15440} {arXiv:2412.15440 [physics.data-an]} \BibitemShut {NoStop}%
\bibitem [{Note1()}]{Note1}%
  \BibitemOpen
  \bibinfo {note} {More detailed benchmarking results, including Apple silicon and other systems, will be provided and kept up-to-date on the GitHub repository \cite {githubarch}.}\BibitemShut {Stop}%
\end{thebibliography}%

\clearpage
\newpage

\appendix
\balance

\section{MC and FNO Training}

\subsection{Hardware and Code}\label{app:hardware}

The \JETSCAPE MC events and FNO training were run on a single machine with an AMD Ryzen Threadripper 3960X Processor,  two NVIDIA GeForce RTX 3090 GPUs (of which we used one at a time), and 128 GB of DDR4 ram. The C++ implementation of the FNOs, and consequent calculations, was run on a local laptop.

Note that the \JETSCAPE code base \cite{Putschke:2019yrg} and \NeuralOperator \cite{kossaifi2025librarylearningneuraloperators} libraries were used in this calculation. Custom code for reading out intermediary \JETSCAPE results, implementing C++ representation of the FNOs, and analysis notebooks for FNO training are all archived at \cite{githubarch}.

\subsection{FNO Training Details}\label{app:training_details}

The following are process details for the custom code generated and choices made in the \texttt{ipynb} analysis files which trained the neural operators.

\begin{enumerate}
    \item A small custom \JETSCAPE model was generated which writes out 
    QGP hydrodynamic evolutions to \texttt{ROOT} files. These are saved
    as in a \texttt{TTree} as \texttt{vector<vector<vector<vector<float>...>},
    in which the indices are, in order, $x$ $(0..59)$, $y$ $(0..59)$, $\tau$
    (variable length until freeze out), and then the three values characterizing the QGP:
    \erho, $v_x$, and $v_y$.
    \item For convenience for reading with \texttt{uproot}, these are converted by a small standalone C++ program to new \texttt{ROOT} files containing fixed length arrays of floats (in place of the nested \texttt{vector} format). This requires choosing that the number of \ttau steps be fixed to a constant, for which 60 is chosen. This value cuts off time steps for most central events, but is too large for most peripheral events, which reach freeze out before the $\tau=\fmc{6.4}$ step (remember that MC forms the QGP at $\tau=\fmc{0.5}$) in more than 97.5\% of events. In the events that are cut-off too soon, the remainder of the array is filled with zeros. This results in an array of size ($60\times60\times60\times3=648,000$) for each event. Any array of much larger than 60 time steps would not fit in the GPU memory for training. The fixed size array allows Python to read the \texttt{ROOT} file data into tensors much more quickly than reading in the nested vectors.
    \item Use the \texttt{uproot} library, in conjunction with the \texttt{awkward} and \texttt{numpy} libraries, to read the input \texttt{ROOT} files. Store the data in a numpy array.
    \item In the generated numpy array, scale the energy density \erho  values by the associated proper time $\tau$ (which ranges from \fmc{0.5} to \fmc{6.4}).
    \item Write custom Python class \texttt{FluidDataSet}, analogous to  \NeuralOperator's \texttt{TensorDataSet} class. This class provides methods to read the numpy input to \texttt{PyTorch} tensor format (in this case on a GPU) for the FNO training.
    \item Use \texttt{FluidDataSet} with \NeuralOperator's \texttt{DataLoader} class to select  the first 80\% of events for training, and the final 20\% events for validation.
    \item Use \NeuralOperator's \texttt{FNO} class for training with \texttt{n\_modes=[30,30,25]} and 64 hidden channels, and a projection-to-channel ratio of 2. Use 3 input and output channels (one each for \erho, $v_x$, and $v_y$).
    \item Train the tensors' first time step ($tau=0.5$) to the following 59 time steps.
    \item Use the AdamW optimzer \cite{loshchilov2019decoupledweightdecayregularization} as implemented in the \NeuralOperator library, with learning rate of 0.008, and weight decay of 0.0004.
    \item Use \NeuralOperator's \texttt{H1Loss} for the training loss, and \texttt{H1Loss} and \texttt{LpLoss} for the evaluation losses. The \texttt{H1Loss} is the Sobolev norm, initialized with 3 dimensions, and the \texttt{LpLoss} is initialized with \texttt{d=3,p=2} to get the second order L-norm (i.e. to evaluate how good the PDEs' derivatives are being modeled).
    \item Train the data using 250 epochs.
    \item Use hooks in \NeuralOperator to automatically save the model training epoch that results in the minimum \texttt{LpLoss}.
\end{enumerate}

\newpage
\onecolumngrid
\section{Additional Figures}

This appending contains additional figures referenced from the body of this paper. Short descriptions of the figure are included here. All figures are from the \JETSCAPE simulations of \sNNcc Au+Au events. The centralities are given in sub- and superscript pairs;  \ie \ICper means the QGP initial conditions in a MC event with centrality set to 40-60\%. The default nucleon width in the simulations is \SI{1.12}{fm}. If the MC uses a different width (either \SI{0.8} or \SI{0.96}), is it listed as a superscript on the left side, \eg \ICcenspikey. 

The results of FNO predictions are labeled according to both MC conditions used to train the FNO (\eg \FNOcen) and the conditions used to generate the IC from which the FNO predicts the QGP evolution (\eg \ICper), with the notation showing the FNO acting on the IC, \eg $\FNOcen(\ICper)$. When compared to the truth-level predictions, calculated by \MUSIC, the comparison is always made for events in which the MC conditions are the same for \MUSIC and the IC. For example, $\FNOcen(\ICper)$ is compared to \MUSICper.

\begin{figure*}[htbp]
    \begin{minipage}{0.48\linewidth}
        \includegraphics[width=0.99\linewidth, trim=1cm 0cm 27cm 1cm, clip]{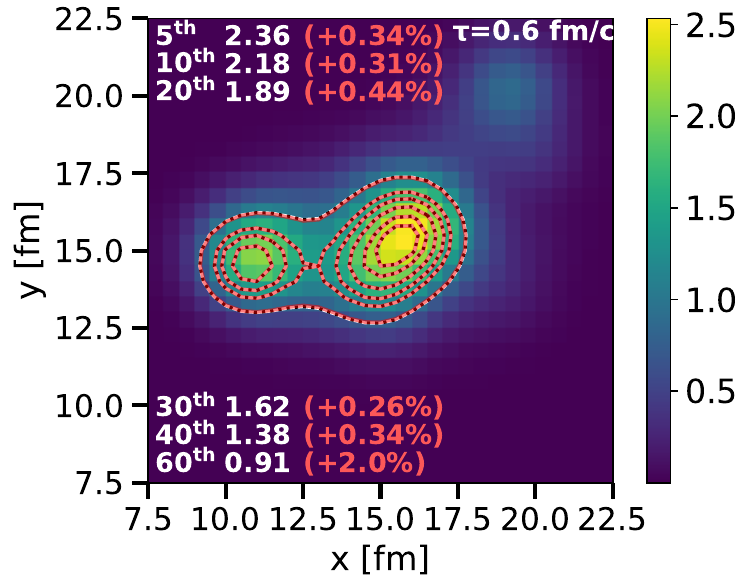} \\
        (a) $\FNOper(\ICper)$, $\tau=\fmc{0.6}$
    \end{minipage}
    \begin{minipage}{0.48\linewidth}
        \includegraphics[width=0.99\linewidth, trim=26.7cm 0cm 1cm 1cm, clip]{periph_8.pdf}\\
        (b) $\FNOper(\ICper)$, $\tau=\fmc{4.4}$
    \end{minipage} \\
    \caption{
    Energy density (\erho) distributions in a peripheral event two \ttau values. The color scale indicates the truth \erho distribution in $\mathrm{GeV}/\mathrm{fm}^3$. The dotted lines show the constant \erho boundaries containing the indicated percentages of the total event energies, with the values of \erho at each line given in white. Lines in red indicate the predicted percentile envelope locations (from $\FNOper(\ICper)$), and values in red indicate the relative constant \erho values of those percentiles relative to the truth.}
    \label{fig:periph_2D}
\end{figure*}

\begin{figure*}[htbp]
    \begin{minipage}{\linewidth}
    \justifying
Fig.~\ref{fig:periph_2D_spikey} is similar to Fig.~\ref{fig:periph_2D}, except that it is a representative event using a smaller nucleon width of \SI{0.8}{fm}. As a result the energy distribution is much sharper, with the hot spot reaching above \GeVcube{6.4} (as opposed to the much smaller value of \GeVcube{2.5} in the event shown in Fig.~\ref{fig:periph_2D}). Consequently, the freeze outs also extend to later times; as seen by the late-time \ttau value being \fmc{5.4} instead of \fmc{4.4} in Fig.~\ref{fig:periph_2D}. The FNO used for the prediction was trained on events miss-matched both the centrality and the nucleon width; \ie $\FNOcen(\ICperspikey)$. The resulting errors are apparent.
    \end{minipage}\\[3em]
    
    \begin{minipage}{0.48\linewidth}
        \includegraphics[width=0.99\linewidth, trim=1cm 0cm 27cm 1cm, clip]{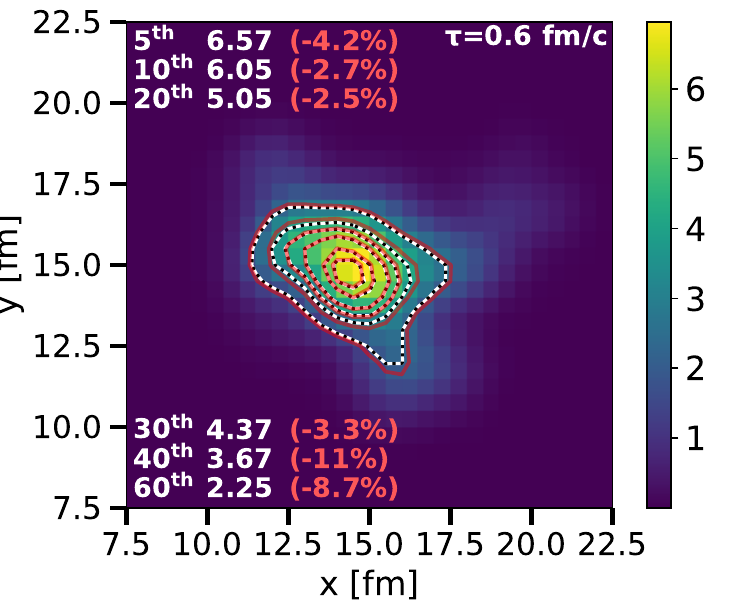}\\
        (a) $\FNOcen(\ICperspikey)$, $\tau=\fmc{0.6}$
    \end{minipage}
    \begin{minipage}{0.48\linewidth}
        \includegraphics[width=0.99\linewidth, trim=26.7cm 0cm 1cm 1cm, clip]{perispiky_5.pdf}\\
        (b) $\FNOcen(\ICperspikey)$, $\tau=\fmc{5.4}$
    \end{minipage} \\
    \caption{
    Energy density (\erho) distributions in a peripheral event with a nucleon width of \SI{0.8}{fm} at two \ttau values. The color scale indicates the truth \erho distribution in $\mathrm{GeV}/\mathrm{fm}^3$. The dotted lines show the constant \erho boundaries containing the indicated percentages of the total event energies, with the values of \erho at each line given in white. Lines in red indicate the predicted percentile envelope locations by an FNO trained on central events with the nominal nucleon width, \ie $\FNOcen(\ICperspikey)$, and values in red indicate the relative constant \erho values of those percentiles relative to the truth.}
    \label{fig:periph_2D_spikey}
\end{figure*}

\begin{figure}[htbp]
    \begin{minipage}{\linewidth}
    \justifying
    Fig.~\ref{fig:cen_on_per} shows late-time \ttau values of a single peripheral event along with predictions of an FNO trained on central events ($\FNOcen(\ICper)$. They differ only the normalization method used. The normalization used in figure (d) is the one used at all other places in this publication.
    \end{minipage}\\[1em]
    
    \centering
    \includegraphics[width=0.9\linewidth, clip, trim=0cm 68cm 36cm 0cm]{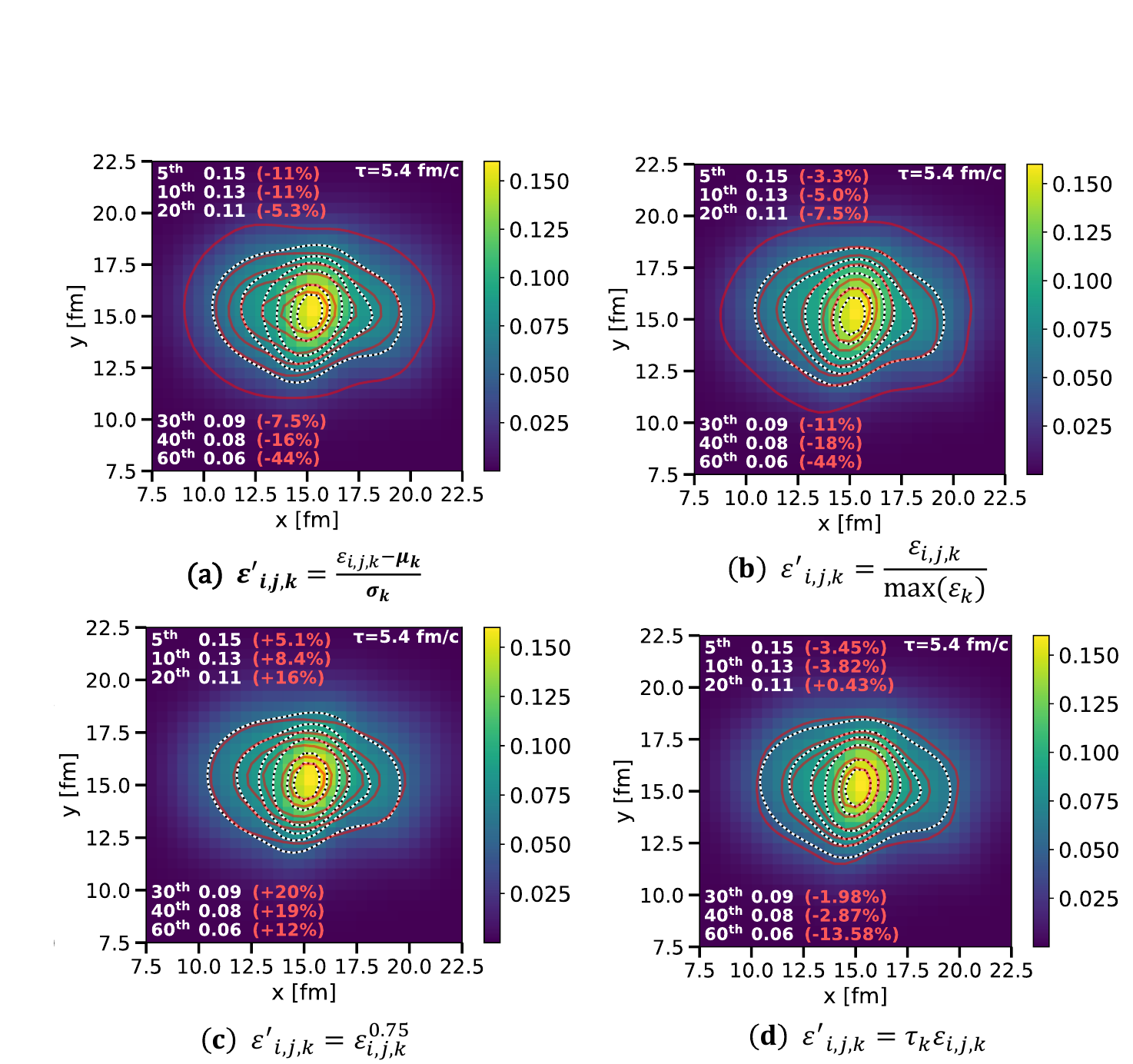}
    \caption{Energy density (\erho) distributions for a \sNNcc Au+Au MC peripheral (40-60\%) collision, with percentile boundaries for total truth energy (in dotted lines with associated \erho values in white text) and FNO predictions (correspondingly in red lines and red text). The FNO in each panel uses an FNO trained with central (0-10\%) collisions using a different data normalization as listed. The normalizations in panels (a) and (b) are commonly used in ML; those in (c) and (d) are inspired by the physics in relativistic hydrodynamic flow.}
    \label{fig:cen_on_per}
\end{figure}

\begin{figure}[htbp]
    \begin{minipage}{\linewidth}
        \justifying
        Fig.~\ref{fig:flow_FNOall_censpikey} shows the \erho distributions at three \ttau steps for a central event with a smaller nucleon width, along with the predictions in an FNO trained on only 500 events of four classes of events: central and peripheral along with nominal and smaller nucleon widths; \ie $\FNOall(\ICcenspikey)$. As shown, the predictions are quite good for the central events at smaller nucleon widths.
    \end{minipage}\\[3em]
    
    \centering
    \includegraphics[width=0.99\linewidth, trim= 1.4cm 0cm 1cm 1cm,clip]{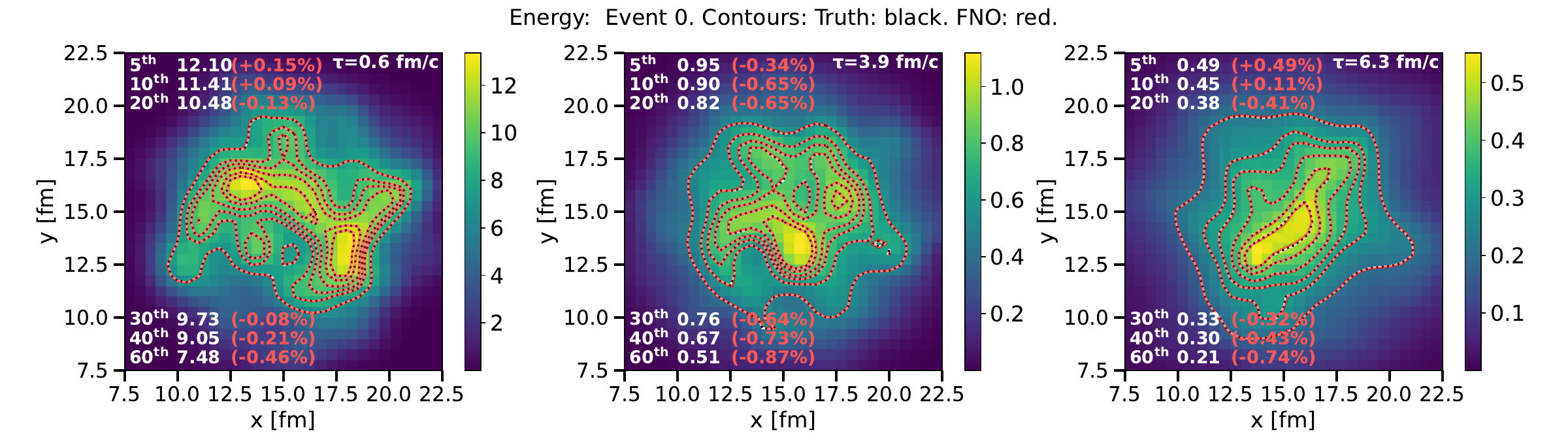} \\
    (a) $\FNOall(\ICcenspikey)$,  \hspace{1.2cm} 
    (b) $\FNOall(\ICcenspikey)$,  \hspace{1.2cm}
    (c) $\FNOall(\ICcenspikey)$,  \\
         $\tau=\fmc{0.6}$ \hspace{4.3cm} 
         $\tau=\fmc{3.9}$ \hspace{4.3cm}
         $\tau=\fmc{6.3}$ \\
    \caption{Energy density \erho distributions' at three \ttau values for a
    central event with a smaller nucleon width in MC at \SI{0.8}{fm}, with
    predictions by an FNO trained on 2,000 ICs: 500 each of \ICcen, \ICper,
    \ICcenspikey, \ICperspikey. For explanation on plotting methodology, see
    caption of Fig.~\ref{fig:periph_2D} above.}
    \label{fig:flow_FNOall_censpikey}
\end{figure}

\begin{figure}[htbp]
    \begin{minipage}{\linewidth}
        \justifying
        Fig.~\ref{fig:erho_halfway} is show similar results to Fig.~\ref{fig:flow_FNOall_censpikey}, using the FNO trained on four sets of ICs (\FNOall), but instead are tested on ICs with nucleon widths between those used to train the FNO (\SI{0.96}{fm}). The results appear equally good to those in Fig.~\ref{fig:flow_FNOall_censpikey}, and indicate that training events with bounding values of nucleon width results in models equally effective for MCs using intermediate nucleon widths. 
    \end{minipage}\\[3em]

    \centering
    \includegraphics[width=0.99\linewidth, trim= 1.4cm 0cm 1cm 1cm,clip]{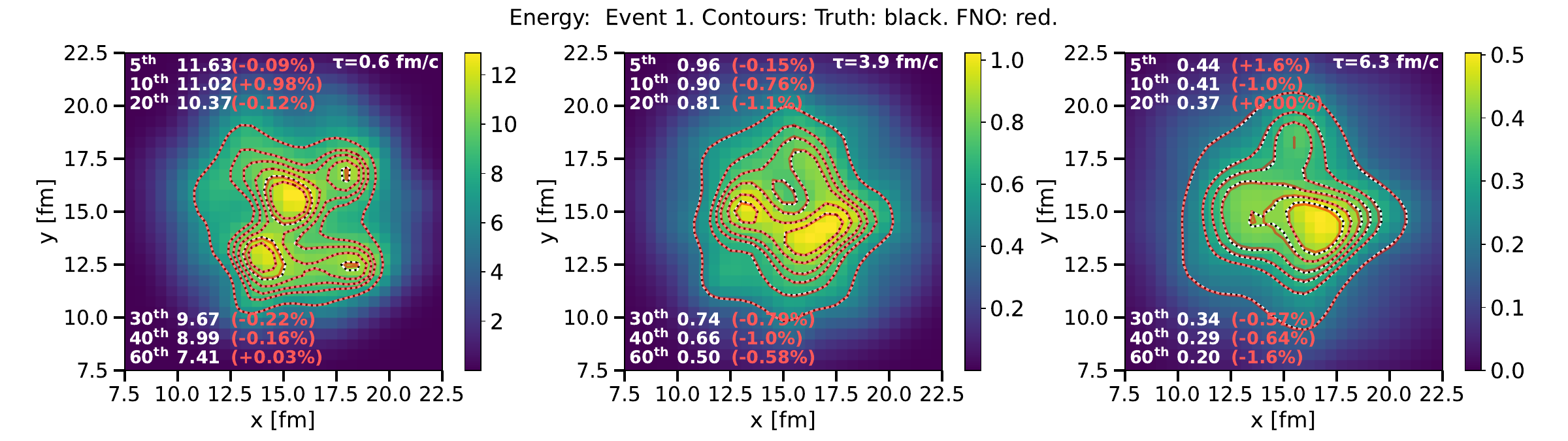} \\
    (a) $\FNOall(\ICcenmidspike)$,  \hspace{1.2cm} 
    (b) $\FNOall(\ICcenmidspike)$,  \hspace{1.2cm}
    (c) $\FNOall(\ICcenmidspike)$,  \\
         $\tau=\fmc{0.6}$ \hspace{4.3cm} 
         $\tau=\fmc{3.9}$ \hspace{4.3cm}
         $\tau=\fmc{6.3}$ \\
    \caption{Energy density \erho distributions' at three \ttau values for a
    central event with IC at a nucleon width of \SI{0.96}{fm}, which is halfway between the difference ICs used to train the \FNOall.
    For explanation on plotting methodology, see
    caption of Fig.~\ref{fig:periph_2D} above.}
    \label{fig:erho_halfway}
\end{figure}

Fig.'s~\ref{fig:periph_2D}-\ref{fig:erho_halfway} show 2D depictions of the energy \xy distribution at different \ttau steps. The color histograms show the \xy distribution of the energy densities \erho in units of $\rm GeV/fm^{3}$. The dotted lines show constant \erho values (analogous to constant elevation lines on a topographical map) bounding percentiles of the total energy in the \xy grid. The values of the \erho for these lines are printed in white. The locations of the lines on the predicted \erho distribution from an FNO, are shown in red, and the percentile errors (relative to the truth \MUSIC values) are printed in white.

Fig.~\ref{fig:periph_2D} shows a typical peripheral event for $\FNOper(\ICper)$. An interesting effect found in the study results from setting all \erho (and \vx and \vy) values past the \ttau's freeze-out time to 0. The trained FNO's learn this, and sometimes in just the final time step or two, the predicted envelop positions of the energy percentiles fluctuate somewhat wildly, although the resulting effect on the hadrons (as shown in this study) is negligible. Also apparent in the figure is the clear increase in azimuthal anisotropy in the initial \erho distribution configuration (leading to a clear \vtwo in the measurement).

\begin{figure}[htbp]
    \begin{minipage}{\linewidth}
        \justifying
        Fig.~\ref{fig:dot_vel} shows the average velocities of the QGP in the \xy plain in the direction radially outward to the central grid-point ($(\SI{15}{fm},\SI{15}{fm})$ in the geometry used in this study). These are shown for a progressing range of radii (also from the central grid point) and \ttau values. The results show excellent agreement of the mean values predicted by the FNO and calculated by \MUSIC.
    \end{minipage}\\[3em]
    \centering
    \includegraphics[width=0.9\linewidth]{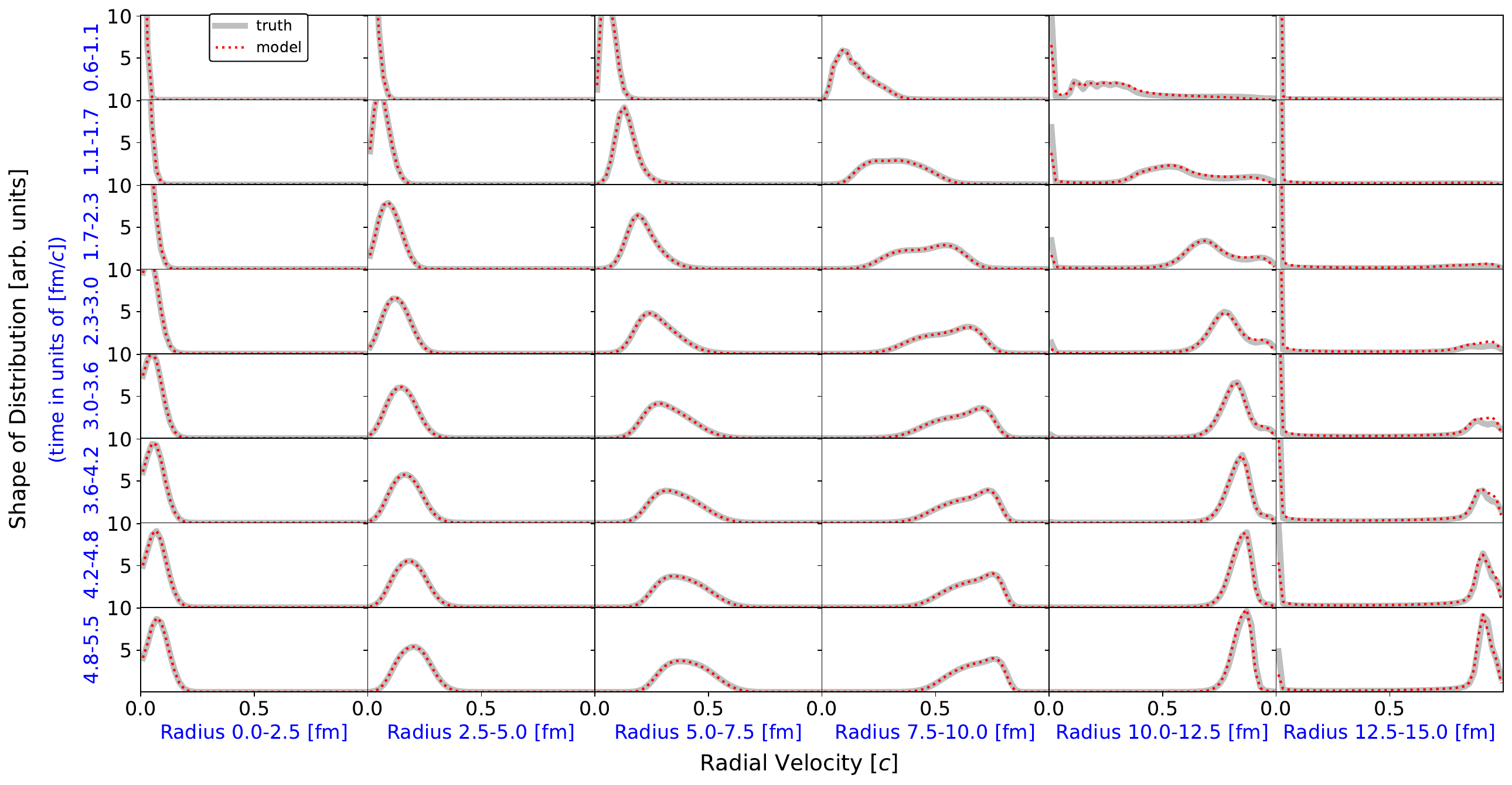}
    \caption{The average radial velocities of 4000 central events with values both from 
    the PDE solver and as predicted by the FNO. In the legend, ``model'' refers to the FNO
    predictions, and ``truth'' to the calculations by \MUSIC.}
    \label{fig:dot_vel}
\end{figure}

\begin{figure}[htbp]
    \begin{minipage}{\linewidth}
        \justifying
        Fig.~\ref{fig:cross_vel} shows the average velocities of the QGP in the \xy plain in the direction normal to those radially outward to the central grid-point ($(\SI{15}{fm},\SI{15}{fm})$ in the geometry used in this study). Positive values are clock-wise and negative values are anti-clock-wise. These are shown for the same sets of radii (also from the central grid point) and \ttau values shown in Fig.~\ref{fig:dot_vel}, and also show excellent agreement of the mean values predicted by the FNO and calculated by \MUSIC.
    \end{minipage}\\[3em]
    \centering
    \includegraphics[width=0.9\linewidth]{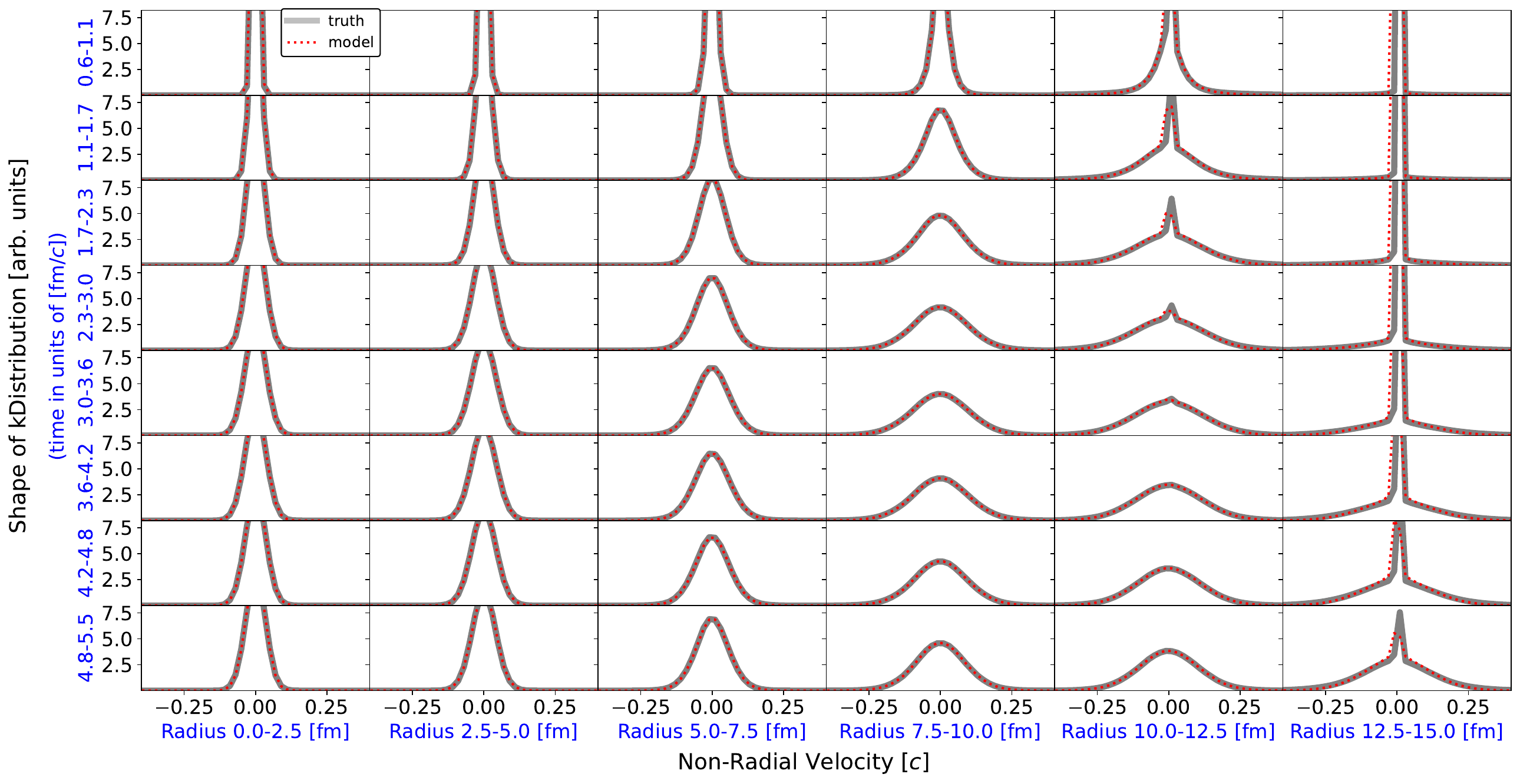}
    \caption{The average values of the azimuthal velocities (i.e. $\vec{v}\times\vec{r}$ where $\vec{r}$
    is the vector to the nominal origin of the MC simulation) of 4000 central
    events. In the legend, ``model'' refers to the FNO predictions, and
    ``truth'' to the calculations by \MUSIC.}
    \label{fig:cross_vel}
\end{figure}

\begin{figure}[htbp]
    \begin{minipage}{\linewidth}
    \justifying
    Fig.~\ref{fig:fnoall_pt} shows the \pT spectra from hadrons from peripheral events at three different nucleon widths as predicted by the FNO trained on two nucleon widths and both centrality classes, and also the truth values calculated by \MUSIC. These are the spectra whose ratios are shown in Fig.~\ref{fig:fnoall_pt_ratio}.
    \end{minipage}\\[1em]
    \begin{minipage}{0.48\linewidth}
    \centering
    \resizebox{1.0\linewidth}{!}{\clipbox{0.3cm 0cm 1.7cm 0cm}{\input{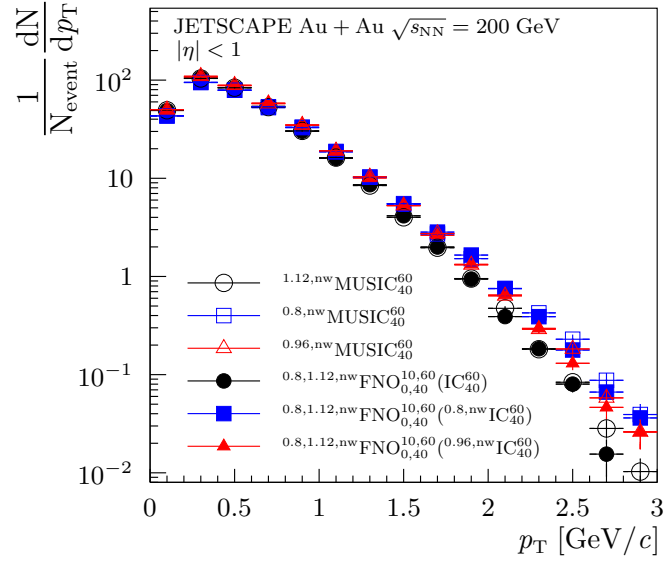}}}
    \begin{tikzpicture}[overlay, remember picture]
        \LabBulkPt
    \end{tikzpicture}
    \vspace{-2em}
    \caption{Hadron \pT spectra of resulting from QGP flow at time of freeze out for three sets of events which are identical except for nucleon width. The results use \FNOall, trained on 500 events each of the largest and smallest nucleon widths with peripheral events (for 1,000 ICs total), and then again for central events (for another 1,000 ICs). The truth spectra are also reported, as calculated by \MUSIC.}
    \label{fig:fnoall_pt}
\end{minipage}
\end{figure}

\begin{figure}[htbp]
    \begin{minipage}{\linewidth}
    \justifying
    Fig.~\ref{fig:jet_zrat_zoomed} is a second view of the three ratios of jet-$z$ (excluding the \MUSIC to $pp$, which indicates the magnitude of the quenching effects at truth level) in Fig.~\ref{fig:jet_z}~(b). The results indicate that, within statistical precision, the FNO predictions are equally accurate with respect to jet quenching as measured by the effects on the $z$-spectra of the jet constituents.
    \end{minipage}\\[1em]
    \begin{minipage}{0.48\linewidth}
        \centering
        \resizebox{1.0\linewidth}{!}{\clipbox{0.3cm 0cm 1.7cm 0cm}{\input{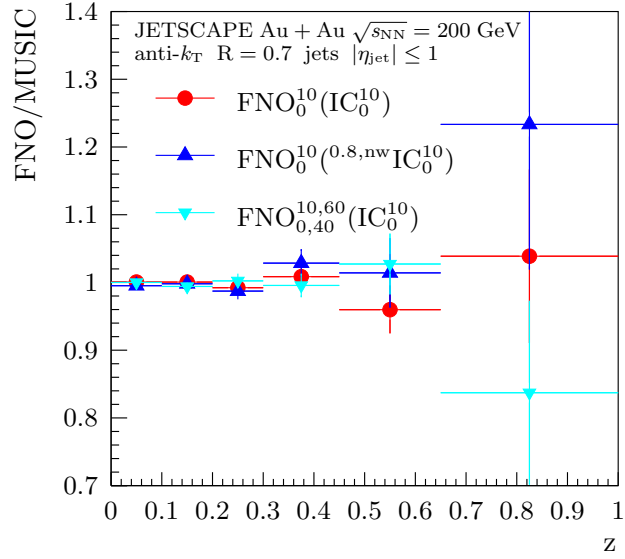}}}
        \begin{tikzpicture}[overlay, remember picture]
            \LabJetZrat
        \end{tikzpicture}
        \vspace{-2em}
        \caption{ Ratios of jet $z$ distributions for jets quenched in QGP modeled by FNOs relative to jets quenched in QGP modeled with \MUSIC (using the same events as the ICs). Integer sub and superscripts indicate the centrality ranges in the MCs used to train the FNO or generate the initial conditions (ICs) of the QGP. The ``0.8 nw'' superscript on the left indicates the \SI{0.8}{fm} nucleon width used for that data; all other use the nominal nucleon width of \SI{1.12}{fm}. (Same as Fig.~\ref{fig:jet_z}~(b) with different y-axis scaling).}
        \label{fig:jet_zrat_zoomed}
    \end{minipage}
\end{figure}

\end{document}